\documentclass[preprints,article,accept,pdftex,moreauthors]{Definitions/mdpi} 
\firstpage{1} 
\pubvolume{1}
\issuenum{1}
\articlenumber{0}
\pubyear{2022}
\copyrightyear{2022}
\datereceived{} 
\dateaccepted{} 
\datepublished{} 
\hreflink{https://doi.org/} 

\usepackage{amsmath,amssymb,amsfonts,physics,graphicx,float,subfiles}
\graphicspath{{pic/}}

\usepackage[labelformat=simple]{subcaption} 

\usepackage{ulem}
\usepackage{color}
\definecolor{ar}{rgb}{1.0, 0.01, 0.24}
\definecolor{al}{rgb}{0.82, 0.1, 0.26}
\definecolor{ev}{rgb}{0.56, 0.0, 1.0}

\newcommand{\PDM}{\mathrm{PDM}}
\newcommand{\CSC}{\mathrm{CSC}}

\newcommand{\hadron}{\mathrm{H}}
\newcommand{\interp}{\mathrm{I}}
\newcommand{\quark}{\mathrm{Q}}

\newcommand{\lag}{\mathcal{L}}
\newcommand{\diag}[1]{\mathrm{diag}(#1)}

\newcommand{\Nf}{N_{\rm f}}

\newcommand{\la}{\langle}
\newcommand{\ra}{\rangle}

\Title{ 
Chiral restoration of nucleons in neutron star matter:\\
studies based on a parity doublet model 
}

\TitleCitation{}

\Author{Takuya Minamikawa$^{1,\ast,\ddagger}$\orcidA{}, Bikai Gao$^{1,\dagger,\ddagger,\P}$, Toru Kojo$^{2,\S,\ddagger}$, and Masayasu Harada$^{1,\P,\ddagger}$}

\AuthorNames{Takuya Minamikawa, Bikai Gao, Toru Kojo, and Masayasu Harada}

\AuthorCitation{Minamikawa, T.; Gao, B.; Kojo, T.; Harada, M.}

\address{%
$^{1}$ \quad Department of Physics, Nagoya University, Nagoya 464-8602, Japan; \\
$^{2}$ \quad Department of Physics, Tohoku University, Sendai 980-8578, Japan: \\
$^\ast$ \quad minamikawa@hken.phys.nagoya-u.ac.jp;\\
$^\dag$ \quad gaobikai@hken.phys.nagoya-u.ac.jp;\\
$^\S$ \quad torujj@nucl.phys.tohoku.ac.jp; \\
$^\P$ \quad harada@hken.pnys.nagoya-u.ac.jp
}

\secondnote{These authors contributed equally to this work.}

\abstract{
We review the chiral variant and invariant components of nucleon masses and its consequence on the chiral restoration in extreme conditions, neutron star matter in particular.
We consider a model of linear realization of chiral symmetry with the nucleon parity doublet structure that permits the chiral invariant mass, $m_0$, for positive and negative parity nucleons.
Nuclear matter is constructed with the parity doublet nucleon model coupled to scalar fields $\sigma$, 
vector fields $(\omega, \rho)$, and to mesons with strangeness through the U(1)$_A$ anomaly. 
In models with large $m_0$, the nucleon mass is insensitive to the medium, 
and the nuclear saturation properties can be reproduced without demanding strong couplings of nucleons to scalar fields $\sigma$ and vector fields $\omega$. 
We confront the resulting nuclear equations of state with nuclear constraints and neutron star observations, and delineate the chiral invariant mass and effective interactions.
To further examine nuclear equations of state beyond the saturation density, we supplement quark models to set the boundary conditions from the high density side.
The quark models are constrained by the two-solar mass conditions, and such constraints are transferred to nuclear models through the causality and thermodynamic stability conditions. 
We also calculate various condensates and matter composition from nuclear to quark matter in a unified matter,
by constructing a generating functional that interpolates nuclear and quark matter with external fields.
Two types of chiral restoration are discussed;
the one due to the positive scalar charges of nucleons, and the other triggered by the evolution of the Dirac sea.
We found the U(1)$_A$ anomaly softens equations of state from low to high density.
}


\begin{document}


\section{Introduction}

Quest for the origin of hadron masses is one of interesting problems in low-energy hadron physics.
Spontaneous chiral symmetry breaking (S$\chi$SB) is known to generate a part of hadron masses.
 A typical model in this context is 
 the linear $\sigma$ model \cite{Schwinger:1957em,Gell-Mann:1960mvl} where the Lagrangian contains nucleons, $N$, and meson fields, $\sigma$, $\vec{\pi}$,
 which are grouped together into a chiral invariant form.
The fields in such models are linear realization of the chiral symmetry which transform linearly under chiral transformations, 
e.g., $N_i \rightarrow N'_i =U_{ij} (\vec{\theta}) N_j$, $\pi_i \rightarrow \pi'_i = V_{i0} (\vec{\theta})\sigma + V_{ij} (\vec{\theta}) \pi_j$ ($\vec{\theta}$: some constant vector).
 The model is arranged to yield the nonzero expectation value of $\sigma$ fields, $\la \sigma \ra$, 
 which breaks the chiral symmetry and generates the nucleon mass term 
 that couples nucleon fields of the left- and right-chirality.
 The nucleon mass ($\propto \la \sigma \ra$) is not chiral invariant, and is entirely generated by the S$\chi$SB.
 
 More general and systematic construction of models is based on the non-linear realization of chiral symmetry 
 for which chiral transformations act on fields non-linearly,
 e.g., $\pi_i \rightarrow U_{ij}(\vec{\pi},\vec{\theta} ) \pi_j $ \cite{Weinberg:1968de}.
 Such pions accompany space-time derivatives which enable us power counting in pion momenta,
 and greatly systematize the construction of effective Lagrangian~\cite{Weinberg:1978kz}.
 The $\sigma$ field does 
not manifestly appear as a dynamical degree
of freedom, and is
 not necessary to make the Lagrangian chiral invariant.
 In fact, we can allow the chiral invariant mass term of the form, $\sim \bar{N} M_{\rm inv} U_5(\vec{\pi}) N = M_{\rm inv} \bar{ {\cal N} } {\cal N}$
 where ${\cal N} \equiv U_5^{1/2} N$ is nucleons (with "pion cloud") in the non-linear realization.
 If we start with a linear $\sigma$ model,
 the chiral invariant mass $M_{\rm inv}$ appear as $\sim ( \la \sigma^2+\vec{\pi}^2 \ra )^{1/2}$,
 but models of non-linear realization do not necessarily require such identification;
this draws our attention to dynamical mechanisms, not necessarily related to the S$\chi$SB, for the origin of $M_{\rm inv}$.

While the non-linear realization allows more general construction of nucleonic models than the linear realization, 
the descriptions without $\sigma$ fields, in practice, have difficulties in the extension to the domain of the chiral symmetry restoration;
there $\vec{\pi}$ and $\sigma$ should together form a chiral multiplet, 
since physical states in symmetry unbroken vacuum must belong to irreducible representations of the chiral symmetry.
We note here that it is not trivial that such mesonic excitations exist in the chiral symmetric phase, but it would be useful to include the $\sigma$ into an effective model to approach the restoration point from the broken phase.  
Furthermore, if 
the chiral restoration is not the 
 first order phase transitions,
one may observe the consequence of the symmetry restoration even before reaching the complete restoration.
For this purpose the linear realization with $\sigma$ has more advantage over the non-linear realization
(where $\sigma$ must be generated dynamically from the pion dynamics).
Such chiral restoration may happen at high temperature and at high density,
and has phenomenological impacts in descriptions of the physics of relativistic heavy-ion collision and neutron stars (NSs) \cite{Lovato:2022vgq}.

A model of linear realization may be improved by supplementing the concept of Weinberg's mended symmetry \cite{Weinberg:1969hw,Weinberg:1990xn}.
The mended symmetry states that, even in spontaneously broken vacuum, 
superposing the linear representations of the original symmetry may be used to describe the physical spectra.
Based on this picture, Weinberg described low-lying mesons $(\sigma, \pi, \rho, a_1)$ as the superposition of chiral multiplets, 
and then obtained reasonable mass relations and decay widths for these states.
This success encourages us to consider models of linear realization for nucleons including several chiral multiplets.

In this review we consider a parity doublet model (PDM) of nucleons as a model of linear realization, 
and examine its feature through the phenomenology of dense QCD, especially neutron star matter.
The PDM includes two nucleon fields, $N_1$ and $N_2$, whose left- and right-handed components (defined through the $(1 \pm \gamma_5)/2$ projections)
transform differently as $N_{1R/L} \rightarrow g_{R/L} N_{1R/L}$ and $N_{2R/L} \rightarrow g_{L/R} N_{2 R/L}$ 
under the U($\Nf$)$_L \otimes $ U($\Nf$)$_R$ chiral transformations (mirror assignment). 
The mass term of $ \sim m_0 \big( \bar{N}_{1R} N_{2L} + \bar{N}_{1L} N_{2R} \big)$ is now possible without breaking U($\Nf$)$_L \otimes $ U($\Nf$)$_R$ symmetry,
and the mass $m_0$ is chiral invariant.
This chiral invariant mass term and the conventional Yukawa coupling term are diagonalized together, yielding spectra of positive and negative parity nucleons.
For a sufficiently large $m_0$, the overall magnitude of the physical nucleon masses is primarily set by $m_0$,
while the chiral variant mass $\propto \la \sigma \ra$ is mainly responsible for the mass splitting between positive and negative parity nucleons.
Such a model was first constructed by DeTar and Kunihiro \cite{Detar:1988kn}, where $N(939)$ and $N^\ast(1535)$ are regarded as partners.

The size of $m_0$ is of great concern to predict the properties of nucleons near the chiral restoration.
In a minimal PDM, the decay width of $N^\ast(1535)$ 
is used to set the constraint $m_0 \lesssim 500$ MeV \cite{Jido:2001nt}.
However, as in the standard $\sigma $ model, such estimates can be easily affected by $\sim 30$\% if we permit non-renormalizable terms of dimension 5, 
and a larger value of $m_0$ is possible (see, e.g., Ref.~\cite{Yamazaki:2018stk}).
Further evidence of large $m_0$ comes from a lattice QCD study at finite temperature for a nucleon and its parity partner~\cite{Aarts:2017rrl}.
The mass gap between $N(939)$ and $N^*(1535)$ is reduced together with the reduction of chiral condensates,
while the substantial mass of $N(939)$ can remain; this suggests that $m_0$ may be as large as the mass of $N(939)$ itself.  

The nucleon mass relatively insensitive to the chiral restoration has important consequences on dense nuclear matter at density relevant for neutron star (NS) phenomenology.
In the past $\sim 20$ years there have been dramatic progress in measurements of NS mass-radius ($M$-$R$) relations which have one-to-one correspondence with the QCD equation of state (EOS).
The key question is whether EOS is stiff or soft; stiffer EOS has a larger pressure at a given energy density and prevents a star from the gravitational collapse to a blackhole.
The relevant NS constraints are 
the existence of $2M_\odot$ NS \cite{Arzoumanian:2017puf,Fonseca:2016tux,Demorest:2010bx,Cromartie:2019kug,Fonseca:2021wxt,Antoniadis:2013pzd},
the radii of $1.4M_\odot$ \cite{TheLIGOScientific:2017qsa,Miller:2019cac,Riley:2019yda}
and $\simeq 2.1M_\odot$ NS \cite{Miller:2021qha,Riley:2021pdl}. 
In short, NS EOS is relatively soft at baryon density $n_B$ around 1-2$n_0$ ($n_0 \simeq 0.16 \,{\rm fm}^{-3}$: nuclear saturation density), but evolves into very stiff EOS at $\sim 5n_0$.
The density $\simeq 1$-$2n_0$ is usually regarded as the domain of nuclear theories, while the domain at $\gtrsim 5n_0$, where nucleons of the radii $\sim 0.5$-$0.8$ fm begin to overlap, likely demands quark matter descriptions.
The EOS constraints at 1-2$n_0$ obviously give important information on the chiral invariant mass, 
but the EOS constraints on $\gtrsim 5n_0$ also impose indirect but powerful constraints on the nuclear territory through the causality condition that the sound velocity, 
$c_s = (\partial P/\partial \varepsilon )^{1/2}$ ($P$: pressure, $\varepsilon$: energy density), is less than the light velocity ($c=1$ in our unit), see, e.g., Ref. \cite{Kojo:2020krb}. 
In order to describe the domain between nuclear and quark matter in a way consistent with the observed soft-to-stiff evolution of EOS,
the simplest scenario is the quark-hadron-crossover (QHC)~\cite{Masuda:2012kf,Masuda:2012ed,Baym:2017whm,Baym:2019iky,Kojo:2021wax}. 
Unlike models with first order phase transitions, gradual quark matter formation does not accompany strong softening of EOS, but even lead to stiffening \cite{Annala:2019puf,Kojo:2021ugu,Kojo:2021hqh,Iida:2022hyy,Brandt:2022hwy}.
Based on this picture,
we build unified equations of state which utilize nuclear models at $n_B\lesssim 2n_0$, quark models at $n_B \gtrsim 5n_0$, and interpolate them for EOS at $2n_0 \lesssim n_B \lesssim 5n_0$.
We confront the unified EOS with $M$-$R$ relations constrained by observations, and also calculate chiral condensates and matter composition.
All these quantities are examined from the nuclear to quark matter domain,
and the correlation between low and high densities gives us global insights into the chiral properties of nucleons.

For construction of nuclear EOS,
we implement a PDM into the Walecka type mean field model with $\sigma$, $\omega$, and $\rho$ \cite{Walecka:1974qa,Serot:1984ey,Serot:1997xg}.
The strangeness is included at the level of $U(1)_A$ anomaly where scalar mesons with strangeness, $\sigma_s$, couple to $\sigma$ made of up- and down-quarks.
For neutron star EOS based on the PDM, see, e.g., Refs.~\cite{Hatsuda:1988mv, Zschiesche:2006zj, Dexheimer:2007tn, Dexheimer:2008cv, Sasaki:2010bp, Sasaki:2011ff,%
Gallas:2011qp, Paeng:2011hy,%
Steinheimer:2011ea,Dexheimer:2012eu, Paeng:2013xya,Benic:2015pia,Motohiro:2015taa,%
Mukherjee:2016nhb,Suenaga:2017wbb,Takeda:2017mrm,Mukherjee:2017jzi,Paeng:2017qvp,%
Marczenko:2017huu,Abuki:2018ijb,Marczenko:2018jui,Marczenko:2019trv,Yamazaki:2019tuo,Harada:2019oaq,Marczenko:2020jma,Harada:2020etl}.
The most notable feature in the PDM is the density dependence.
The chiral invariant mass allows nucleons to stay massive during the reduction of $\la \sigma \ra$.
In dilute regime the $\la \sigma \ra$ decreases linearly as a function of $n_B$, and so does the nucleon mass $\propto \la \sigma \ra$ if $m_0$ is absent;
the nucleon mass at $n_0$ is $\simeq 30$-50\% smaller than the vacuum mass.
With a larger $m_0$, the mass reduction becomes more modest.
In addition, nucleon fields need not to couple to $\sigma$ very strongly to reproduce the nucleon mass $m_N \simeq 939$ MeV;
in the Walecka type model, this results in a weaker coupling between nucleons and $\omega$ fields,
because such models have been arranged to balance the attractive $\sigma$ and repulsive $\omega$ contributions to reproduce the nuclear matter properties at $n_0$.
Beyond $n_0$, the attractive $\sigma$ contributions decrease while the repulsive $\omega$ contributions keep growing.
Thus, a greater $m_0$ makes the overall magnitude of $\sigma$ and $\omega$ contributions smaller, 
and the resulting softer $\omega$ repulsion improves the consistency with the radius constraints on $1.4M_{\odot}$ NS
for which EOS at $n_B=1$-$2n_0$ is most important.

The PDM as a hadronic model does not describe the chiral restoration at quark level, such as the modification in the quark Dirac sea.
In order to supply such qualitative trend, quark matter EOS plays a role as a high density boundary condition.
For the quark matter, a three flavor Nambu--Jona-Lasinio (NJL)-type model, 
which leads to the color-flavor locked (CFL) color-superconducting matter (for a review, Ref.\cite{Alford:2007xm}), is adopted.
 Effective interactions are examined to fulfill the two-solar-mass ($2M_\odot$) constraint \cite{Kojo:2014rca,Baym:2019iky,Song:2019qoh,Kojo:2021wax}.
In Refs.~\cite{Marczenko:2019trv,Marczenko:2020jma}, 
they construct an effective model combining a PDM and
an NJL-type model with two flavors assuming no color-superconductivity.

This article is mostly a review of our works Refs.\cite{Minamikawa:2020jfj},~\cite{Gao:2022klm}, and \cite{Minamikawa:2021fln}, 
but also presents improved the analyses of Refs.\cite{Minamikawa:2020jfj} and \cite{Minamikawa:2021fln} with the up-to-date version of our PDM.
In Ref. \cite{Minamikawa:2020jfj}, we used the PDM without $U(1)_A$ anomaly to construct a unified EOS, 
and obtained the constraint $600\,\mbox{MeV} \lesssim m_0 \lesssim 900\,\mbox{MeV}$.
The lower bound is primarily determined by the tidal deformability constraint from the GW170817, a detection of gravitational waves from a NS merger event.
Later, in Ref.~\cite{Gao:2022klm}, we updated the PDM by adding the $U(1)_A$ anomaly effects, 
the Kobayashi-Maskawa-'t\,Hooft (KMT) interactions \cite{Kobayashi:1970ji}, to the meson sector.
Even though we stop using the PDM to $\lesssim 2n_0$ before hyperons appear,
the strangeness does affect the chiral condensates in up- and down-sectors through the KMT interactions.
The $U(1)_A$ effects enhance the energy difference between chiral symmetric and broken vacua,
leading to stronger softening in EOS when the chiral symmetry is restored. 
This is found to be true for both hadronic and quark matter.
Especially,  the chiral restoration with the $U(1)_A$ anomaly makes EOS at 1-$2n_0$ softer and leads to small radii for $1.4M_\odot$ NS.
In effect, the lower bound $m_0 \gtrsim 600$ MeV given in Ref.\cite{Minamikawa:2020jfj} is relaxed to $m_0 \gtrsim 400$ MeV.
%
%

While seminal works \cite{Masuda:2012kf,Masuda:2012ed,Masuda:2015kha,Kojo:2014rca,Fukushima:2015bda,Baym:2019iky,Minamikawa:2020jfj,Komoltsev:2021jzg}
utilize the interpolation to construct unified EOS,
microscopic quantities have not been calculated in a unified way.
To utilize the full potential of the interpolation framework,
in Ref.~\cite{Minamikawa:2021fln}
three of the present authors (TM, TK, MH) extended the interpolation to unified generating functionals with external fields coupled to the quantities of interest,
and differentiated the functionals to extract chiral and diquark condensates as well as matter composition.
%
The condensates in the interpolated domain are affected by the physics of both hadronic and quark matter through the boundary conditions for the interpolation;
 for $m_0 \gtrsim 500$\,MeV the significant chiral condensate remains to $n_B \sim 2$-$3n_0$, and  
smoothly approach the condensate in the quark matter at $n_B \gtrsim 5n_0$. 
In this review, we update these analyses including the effects of the $U(1)_A$ anomaly.

This review is structured as follows.
In Sec.~\ref{sec:PDM},
we first review the PDM with mesonic potentials in Ref.~\cite{Gao:2022klm}, 
and show how to constrain the model parameters to satisfy  the hadron properties at vacuum and the saturation properties in nuclear matter.
Sec.~\ref{sec:QM NJL} is the review of quark matter construction.
With these hadronic and quark matter models, in Sec.~\ref{sec:interpolation} we construct unified generating functionals as introduced in Ref.~\cite{Minamikawa:2021fln}, 
and calculate various condensates.
Sec.~\ref{sec:summary} is devoted to summary.


\section{Hadronic matter from a parity doublet model}
\label{sec:PDM}

In this section, 
we review the construction of the PDM in Ref.~\cite{Gao:2022klm}.
The fields appearing in the Lagrangian are linear realization of chiral symmetry,
classified by the chiral representation as (SU(3)$_L$, SU(3)$_R$)$_{{\rm U(1)}_A}$.
We determine the model parameters to reproduce hadronic properties in vacuum and the saturation properties of nuclear matter.

\subsection{Scalar and pseudoscalar mesons}

We introduce a $3\times3$ matrix field $\Phi$ for scalar and pseudoscalar mesons 
which belong to $(\mathbf{3},\mathbf{\bar{3}})_{-2}$ under (SU(3)$_L$, SU(3)$_R$)$_{{\rm U(1)}_A}$ symmetry.
The Lagrangian is given by
%
\begin{align}
\mathcal{L}_{M}^{\rm{scalar}}=\frac{1}{4} \operatorname{tr}\left[\partial_{\mu} \Phi \partial^{\mu} \Phi^{\dagger} \right]-V_{M}-V_{\rm{SB}} - V_{{\rm Anom}},
\end{align}
where 
%
%
\begin{align}
V_{M}=&-\frac{1}{4} \bar{\mu}^{2} \operatorname{tr}\left[\Phi \Phi^{\dagger}\right]+\frac{1}{8} \lambda_{4} \operatorname{tr}\left[\left(\Phi \Phi^{\dagger}\right)^{2}\right] 
-\frac{1}{12} \lambda_{6} \operatorname{tr}\left[\left(\Phi \Phi^{\dagger}\right)^{3}\right] \notag\\
& +\lambda_{8} \operatorname{tr}\left[\left(\Phi \Phi^{\dagger}\right)^{4}\right] 
+\lambda_{10} \operatorname{tr}\left[\left(\Phi \Phi^{\dagger}\right)^{5}\right] ,\\
V_{\rm{SB}}=&-\frac{1}{2}c \operatorname{tr}
	\left[ \hat{m}_q^{\dagger} \Phi + \hat{m}_q \Phi^{\dagger}\right],\\
V_{\rm{Anom}}=&- B\left[\operatorname{det}(\Phi)+\operatorname{det}\left(\Phi^{\dagger}\right)\right]\,,
\end{align}
with $B$ and $c$ being the coefficients for the axial anomaly term and the explicit chiral symmetry breaking term, respectively, 
and 
$\hat{m}_q ={\rm{diag}}\{m_{u}, m_{d}, m_{s}\}$. 
In the above, we include only terms with one trace in $V_{M}$ since they are of leading order in the $1/N_{c}$ expansion.

The present hadronic model is used up to $2 n_0$ assuming no appearance of hyperons.
In the mean field approximation, the $\Phi$ field can be reduced to
\begin{align}
    \Phi \rightarrow
    \left(
    \begin{matrix}
    M&0\\
    0&\sigma_{s}
    \end{matrix}
    \right)_{3\times 3} \,,
\end{align}
where $M$ is a $2\times2$ matrix field transforming as $M \rightarrow g_{L}M g_{R}^{\dagger}$ with
$g_{L,R} \in \mbox{SU(2)}_{L,R}$.
While we apply the mean field, here we keep a matrix representation for the two-flavor part
to clarify the symmetry of the two-flavor part.
The field $\sigma_s$ corresponds to the 
scalar condensate made of a strange and an anti-strange quarks, $\left\langle \bar{s} s \right\rangle$.
The reduced Lagrangian is now given by
\begin{align}
\mathcal{L}_{M}^{\rm{scalar}}
=\frac{1}{4}
\bigg( {\rm tr} \left[ \partial_{\mu}M \partial^\mu M^{\dagger} \right]
+ ( \partial_{\mu} \sigma_{s}\partial^{\mu} \sigma_s) 
\bigg)
- V_M - V_{SB} - V_{\rm Anom} \, ,
\end{align}
where
\begin{align}
V_{M}=&-\frac{1}{4} \bar{\mu}^{2} \left( {\rm tr} \left[MM^{\dagger} \right] 
+ \sigma_{s}^2 \right)
+\frac{1}{8} \lambda_{4} \left( {\rm tr} \left[(MM^{\dagger})^{2}\right] +\sigma_s^{4}\right)\nonumber\\
&{}-\frac{1}{12} \lambda_{6} \left({\rm tr} \left[ (MM^{\dagger})^{3} \right]+ \sigma_s^6 \right)
+\lambda_{8} \left( {\rm tr} \left[(MM^{\dagger})^{4}\right] + \sigma_{s}^8\right)\nonumber\\
&+\lambda_{10}\left({\rm tr} \left[(MM^{\dagger})^{5}\right] + \sigma_s^{10}\right) ,\\
V_{\rm{S B}}=&-\frac{\, c \,}{2} 
\bigg[
{\rm tr} \left[ \hat{m}_{2\times2}
(M+M^{\dagger})\right]+ 2 m_{s} \sigma_s  \bigg], 
\label{VSB term}\\
V_{\rm{Anom}}=&- B \sigma_s \left[ {\rm{det}}(M) + {\rm det(}M^{\dagger}) \right]\,,
\label{V anom}
\end{align}
%
with $\hat{m}_{2\times2}={\rm diag}\{m_{u},m_{d}\}$. 
In the mean field treatment, the two-flavor matrix field $M$ is reduced to ${\rm diag}(\sigma, \sigma)$.

\begin{figure}[htp]
\centering
\includegraphics[width=7.5cm]{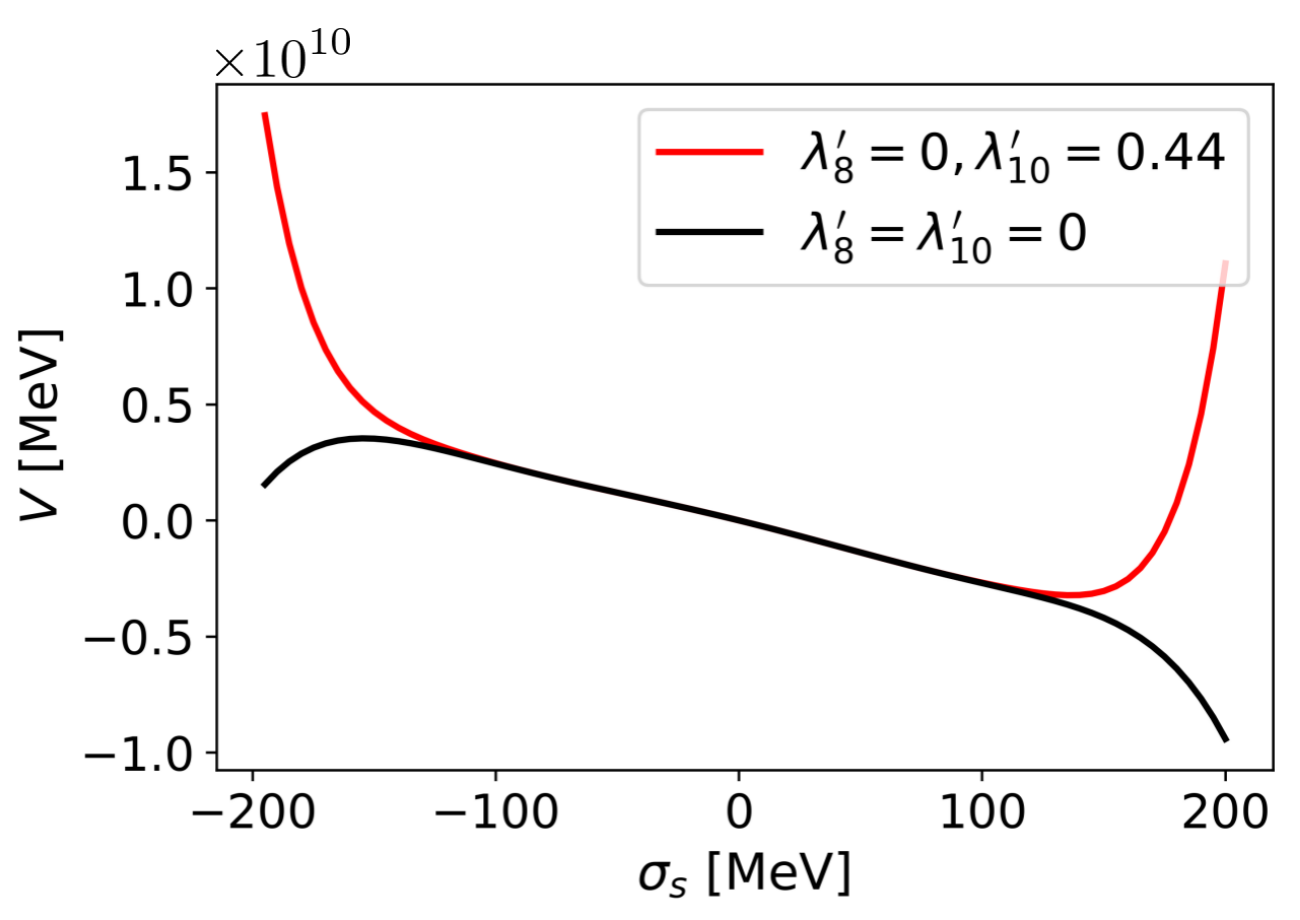}
\caption{
Potential for $\sigma_{s}$ in the vacuum with $m_{0}=800$ MeV.
Here $\lambda'_8$ and $\lambda'_{10}$ are dimensionless quantities defined by 
$\lambda_{8}^{\prime} = \lambda_{8} f_{\pi}^{4}$ and $\lambda_{10}^{\prime} = \lambda_{10}f_{\pi}^{6}$. 
}
\label{V_sigmas}
\end{figure}

Here one might wonder why the $(\Phi \Phi^\dag)$ terms are included up to the fifth powers.
In fact, the potential of the two-flavor model in the analyses~\cite{Motohiro:2015taa,Yamazaki:2019tuo,Minamikawa:2020jfj}
is not bounded from below at a very large value of $(\Phi \Phi^\dag)$, but has only a local minimum.
There,
very large values of $(\Phi \Phi^\dag)$ are simply discarded because they are not supposed to be within the domain applicability of the model.
For three-flavor models with the KMT interactions, however, it turns out that reasonable local minima do not exist as seen by the black curve in Fig.~\ref{V_sigmas}. 
We add higher order terms to stabilize the potential and fine tune models to reproduce the nuclear saturation properties.
Note that these higher order terms do not modify the potentials at small $\sigma_s$.
%
%
%


\subsection{ Nucleon parity doublet and vector mesons   }

In the analysis done in Refs.~\cite{Minamikawa:2020jfj,Minamikawa:2021fln,Gao:2022klm}, hadronic models are used only up to $2n_0$ with assuming that hyperons are not populated.
So, although the mesonic sector includes three-flavors, we include only nucleons in the baryon sector.
The 
nucleons and the chiral partners 
 belong to the $(\mathbf{2},\mathbf{1})_{+1}$ 
and $ (\mathbf{1}, \mathbf{2})_{-1}$  representations under $(\mbox{SU(2)}_L\,,\,\mbox{SU(2)}_R)_{{\rm U(1)}_A}$:
\begin{align}
    \psi_1^{L}: (\mathbf{2},\mathbf{1})_{-1},\quad \psi_1^{R}: (\mathbf{1},\mathbf{2})_{+1}, \quad
    \psi_2^{L}: (\mathbf{1},\mathbf{2})_{+1},\quad \psi_2^{R}: (\mathbf{2},\mathbf{1})_{-1},
\end{align}
%
We note that $\psi_1$ and $\psi_2$ carry the positive and negative parities, respectively:
\begin{align}
\psi_1 \ \mathop{\rightarrow}_{P} \ \gamma_0 \psi_1 \ , \quad \psi_2 \ \mathop{\rightarrow}_{P} \ - \gamma_0 \psi_2 \ .
\end{align}
The relevant  
Lagrangian for nucleons and their Yukawa interactions to the field $M$ is given by
\begin{align}
\lag_N^{\rm scalar}
= & \sum_{i=1,2} \bar{\psi}_{i} i \gamma^{\mu} D_{\mu} \psi_{i} -m_{0}\left(\bar{\psi}_{1}^{L} \psi_{2}^{R}-\bar{\psi}_{1}^{R} \psi_{2}^{L}-\bar{\psi}_{2}^{L} \psi_{1}^{R}+\bar{\psi}_{2}^{R} \psi_{1}^{L}\right)\nonumber\\
&{} -g_{1}\left(\bar{\psi}_{1}^{L}\tau^{2} (M^{\dagger})^{\rm{T}}\tau^{2} \psi_{1}^{R}+\bar{\psi}_{1}^{R} \tau^{2} M^{\rm{T}} \tau^{2} \psi_{1}^{L}\right)\nonumber\\
&{} 
-g_{2}\left(\bar{\psi}_{2}^{L} \tau^{2} M^{\rm{T}} \tau^{2} \psi_{2}^{R}+\bar{\psi}_{2}^{R} \tau^{2} (M^{\dagger})^{\rm{T}}\tau^{2} \psi_{2}^{L}\right)  \ ,
\label{Lag N scalar}
\end{align}
where the covariant derivatives on the nucleon fields are defined as
\begin{align}
D_\mu\psi_{1,2}=(\partial_\mu-i V_\mu)\psi_{1,2}\,, \end{align}
with
\begin{align}
	V_\mu =
	\begin{pmatrix} \mu_B + \mu_Q & 0 \\ 0 & \mu_B \end{pmatrix} \delta_\mu^0 \ .
\end{align}
%
Following Ref.~\cite{Motohiro:2015taa}, the vector mesons $\omega$ and $\rho$ are included based on the framework of the hidden local symmetry (HLS)~\cite{Bando:1987br,Harada:2003jx}.
Here 
instead of showing the forms manifestly invariant under the HLS, 
we just show the relevant interaction terms 
among baryons and vector mesons:
\begin{align}
{\mathcal L}_{N}^{\rm vector} = & - \sum_{i=1,2} \, \bar{\psi}_i \,\gamma^\mu \left( g_{\omega NN} \omega_\mu + g_{\rho NN} \,\frac{\vec{\tau}}{2} \, \vec{\rho}_\mu \right) \psi_i \ , 
\end{align}
where $\vec{\tau}$ is the Pauli matrix for iso-spin symmetry.
The relevant potential terms for vector mesons are expressed as
\begin{align}
{\mathcal L}_{V} = & \frac{1}{2}\, m_\omega^2 \, \omega_\mu \omega^\mu + \frac{1}{2} \, m_\rho^2 \, \vec{\rho}_\mu \cdot \vec{\rho}^\mu + \lambda_{\omega\rho} g_\omega^2 g_\rho^2 \, \omega_\mu \omega^\mu\, \vec{\rho}_\nu \cdot \vec{\rho}^\nu \, .
\label{vector lag}
\end{align}
In the presence of $\omega$,
the attractive $\omega^2$-$\rho^2$ coupling with $\lambda_{\omega\rho} >0$ assists the appearance of $\rho$ fields,
reducing the symmetry energy associated with the isospin asymmetry, as discussed below.
(Note that $\lambda_{\omega\rho}>0$ is needed for the VEVs of $\omega$ and $\rho$ fields not to have non-zero value at vacuum.)

In the mean field approximation, the meson fields take
\begin{align}
	\expval{\sigma}&=\sigma \ , \quad 
	\expval{\omega^\mu}=\omega\delta_0^\mu \ , \quad 
	\expval{\rho^\mu}=\rho \frac{\tau_3}{2} \, \delta_0^\mu \ ,
\end{align}
where each mean field is assumed to be independent of the spatial coordinates. 
The thermodynamic potential in the hadronic matter is calculated as~\cite{Motohiro:2015taa} 
\begin{align}
\Omega_\PDM
=&V(\sigma, \sigma_s)-V(\sigma_0, \sigma_{s0})-\frac12m_\omega^2\omega^2-\frac12m_\rho^2\rho^2 \nonumber \\
&{}-\lambda_{\omega \rho}\left(g_{\omega} \omega\right)^{2}\left(g_{\rho} \rho\right)^{2}
	-2\sum_{i=\pm}\sum_{\alpha=p,n}\int^{k_F^{\alpha,i}}_{\mathbf{p}} 	(\mu_\alpha^*-E^i_\mathbf{p}) \ ,\label{PDM: grand functional}
\end{align}
where $i=+ $ and $-$ label the ordinary nucleon $N(939)$ and the excited nucleon $N^*(1535)$,  respectively.
The energies of these nucleons are $E_\mathbf{p}^i=\sqrt{\mathbf{p}^2+m_i^2}$ with the momenta $\mathbf{p}$
and masses obtained by diagonalizing the Lagrangian (\ref{Lag N scalar}),
\begin{align}
m_\pm = \sqrt{ m_0^2 + \left(\frac{g_1 + g_2}{2} \, \sigma \right)^2 } \mp \frac{g_2 - g_1}{2}\, \sigma \ , \quad
\label{N masses}
\end{align}
where $g_2 > g_1$ is assumed so that $m_+ < m_-$.
The effective chemical potentials $\mu_p^\ast$ and $\mu_n^\ast$ are defined as
\begin{align}
\mu_p^\ast=\mu_B+\mu_Q-g_{\omega NN}\, \omega-\frac{1}{2}g_{\rho NN}\,\rho \ , \quad
\mu_n^\ast=\mu_B-g_{\omega NN}\, \omega+\frac{1}{2}g_{\rho NN}\,\rho \ ,
\end{align}
In the integration above, the integral region is restricted as $|\mathbf{p}|<k_F^{\alpha,i}$ where 
$k_F^{\alpha,i} =\sqrt{(\mu_\alpha^\ast)^2-m_i^2}$ is the Fermi momentum for a nucleon $i$.
In the above expression, we implicitly used the so called no sea approximation, 
assuming that the structure of the Dirac sea remains the same for the vacuum and medium for $n_B \lesssim 2n_0$.
$V(\sigma,\sigma_s)$ is the potential of scalar mean fields given by 
\begin{align}
 V(\sigma,\sigma_{s})=&-\frac{1}{2} \bar{\mu}^{2}\left(\sigma^{2}+\frac{1}{2} \sigma_{s}^{2}\right)+\frac{1}{4} \lambda_{4}\left(\sigma^{4}+\frac{1}{2} \sigma_{s}^{4}\right) -\frac{1}{6} \lambda_{6}\left(\sigma^{6}+\frac{1}{2} \sigma_{s}^{6}\right)\nonumber\\
 &+\lambda_{8}\left(2 \sigma^{8}+\sigma_{s}^{8}\right) +\lambda_{10}\left(2 \sigma^{10}+\sigma_{s}^{10}\right)-2 B \sigma^{2} \sigma_{s} 
-\left(2 c m_{u} \sigma+c m_{s} \sigma_{s}\right) \,.
\label{mean field potential}
\end{align}
In Eq.~(\ref{PDM: grand functional}) we subtracted the potential in vacuum $V(\sigma_0,\sigma_{s0})$, with which the total potential in vacuum is zero.
Here $\sigma_0$ and $\sigma_{s0}$ are related with the decay constants $f_\pi$ and $f_K$ as
\begin{align}
\sigma_0 = f_\pi \ , \quad \sigma_{s0}= 2 f_K - f_\pi \ .
\end{align}

Finally, we include leptons for the charge neutrality realized in NSs.
The total thermodynamic potential of hadronic matter for NSs takes the form
\begin{align}
	\Omega_\hadron=\Omega_\PDM+\sum_{l=e,\mu}\Omega_l \ ,
\end{align}
where $\Omega_l$ ($l = e,\mu$) are 
the thermodynamic potentials for leptons given by
\begin{align}\label{pot: lepton}
\Omega_l=-2\int^{k_F^l}_{\mathbf{p}}(\mu_l-E_\mathbf{p}^l) \,,
~~~~~~~~~~~
k_F^{l} =\sqrt{ \mu_l^2 - m_l^2 } \,.
\end{align}
Here, the mean fields are determined by the following stationary conditions:
\begin{align}\label{gap: PDM}
0=\pdv{\Omega_\hadron}{\sigma}
=\pdv{\Omega_\hadron}{\omega}
=\pdv{\Omega_\hadron}{\rho}\,. 
\end{align}
In neutron stars, 
we impose the beta equilibrium and the charge neutrality condition represented as
\begin{align}
\label{beta eq.}	
\mu_e&=\mu_\mu=-\mu_Q \ , \\
\pdv{\Omega_\hadron}{\mu_Q}&=n_p-n_l=0 \ .
\end{align}
The mean fields and charge chemical potential is determined as functions of $\mu_B$.
After substituting these values into $\Omega_\hadron$, 
we obtain the pressure in the hadronic matter as a function of $\mu_B$,
\begin{equation}\label{P_H}
P_\hadron (\mu_B) = - \Omega_\hadron (\mu_B) \ .
\end{equation}

\subsection{Determination of model parameters}

\label{sec:parameters}

\begin{figure}[htp]
\centering
\includegraphics[width=7.5cm]{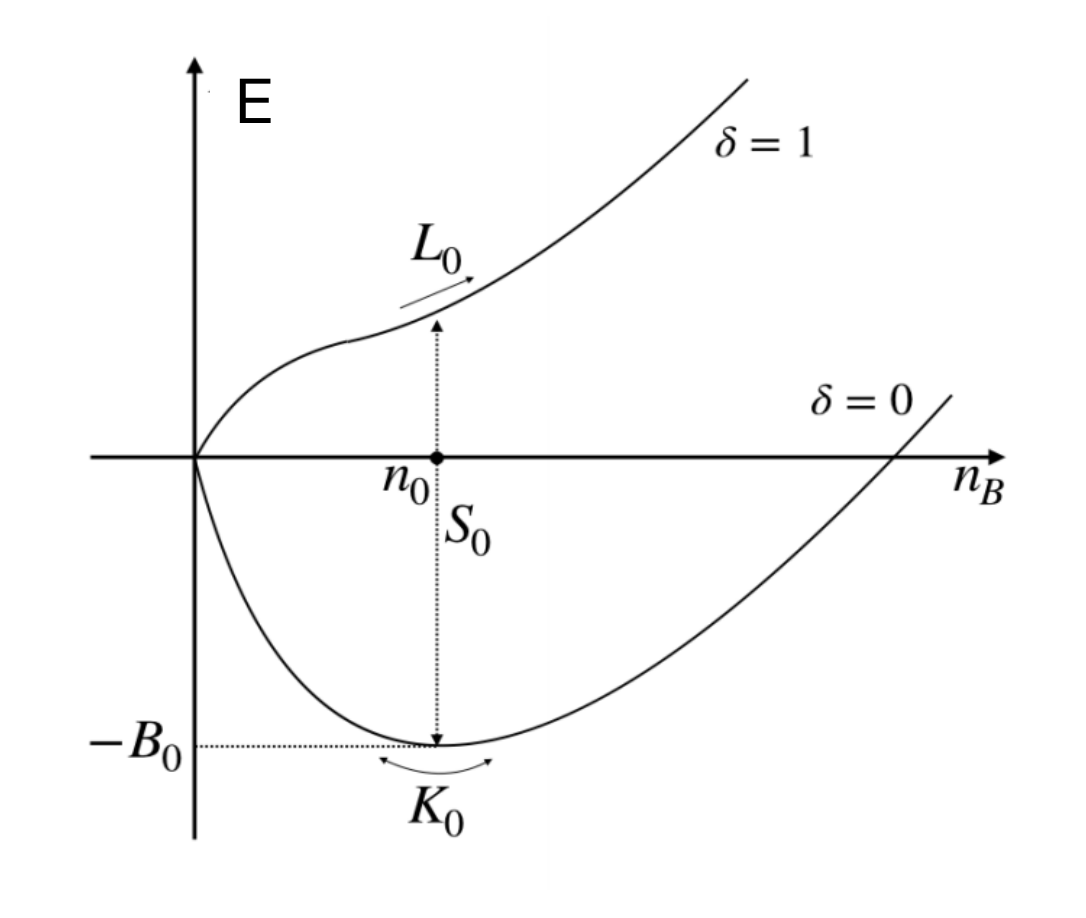}
\caption{
Density dependence of energy per nucleon for the symmetric matter (indicated by $\delta=0$) 
and the pure neutron matter ($\delta=1$). 
}
\label{saturation_fig}
\end{figure}

In this subsection, we determine the parameters in the  PDM 
to reproduce the masses and decay constants in vacuum 
and the saturation properties in nuclear matter. 
In nuclear matter, 
the energy per nucleon (energy density) $\varepsilon$ is given as 
a function of the baryon number density $n_B$ and the isospin asymmetry $\delta = \frac{ n_{p} - n_{n} }{n_{p} + n_{n}}$. 
The energy density is expanded around the normal nuclear density $n_{0}=0.16\, {\rm fm}^{-3}$ and symmetric matter $\delta=0$ as
\begin{align}
\varepsilon = - B_{0} + \frac{K_{0}}{2}\left( \frac{\, n_{B}-n_{0} \,}{3n_{0}}\right)^{2} 
	+  \delta^{2} \bigg( S_{0} + L_{0}\, \frac{\, n_{B}-n_{0}}{3n_{0} \,} \bigg) + {\rm higher\,order},
\end{align}
here the $B_{0}, K_{0}, S_{0}, L_{0}$ are called the binding energy, incompressibility, symmetry energy and slope parameter, respectively, as shown in Fig.~\ref{saturation_fig}.  
The parameter $K_{0}$ measures the curvature of the energy density at the normal nuclear density:
\begin{align}
K_{0} = 9 n_{0}^{2} \, \frac{\partial^{2} \varepsilon}{\, \partial n_{B}^{2} \,} \bigg|_{ n_{B}=n_{0}, \, \delta=0 }.
\end{align}
The symmetry energy $S_{0}$ is calculated as
\begin{align}
S_{0} = \frac{1}{\,2 \,}  \,\frac{\partial^2 \varepsilon}{\, \partial \delta^2 \,} \bigg\vert_{n_B = n_0,  \, \delta = 0} \ .
\end{align}
The parameter $L_0$  characterize the slope of symmetry energy at normal nuclear density:
\begin{align}
L = \frac{3n_0}{2} \, \frac{\partial^3 \varepsilon}{\, \partial n_B \, \partial \delta^2 \,}  \bigg\vert_{n_B = n_0, \, \delta = 0} \ .
\end{align}
We summarize input values which we used in this review in Tables.~\ref{input: mass} and \ref{saturation}. 
%
\begin{table} [htb]
\centering
	\caption{  {\small Physical inputs in vacuum in unit of MeV.  }  }\label{input: mass}
	\begin{tabular}{cccccccc}
		\hline\hline
		~$m_\pi$~&~ $m_K$ ~&~ $f_\pi$ ~&~ $f_{K}$ ~&~ $m_\omega$ ~&~ $m_\rho$ ~&~ $m_+$ ~&~ $m_-$\\
		\hline
		~140 ~&~494~&~ 92.4 ~&~ 109 ~&~ 783 ~&~ 776 ~&~ 939 ~&~ 1535\\
		\hline\hline
	\end{tabular}
	\vspace{0.3cm}
\end{table}
\begin{table} [htb]
\centering
	\caption{  {\small Saturation properties used to determine the model parameters: the saturation density $n_0$, the binding energy $B_0$, the incompressibility $K_0$, symmetry energy $S_0$ and the slope parameter $L_{0}$.}  
	}
	\begin{tabular}{ccccc}\hline\hline
	~$n_0$ [fm$^{-3}$] ~& $E_{\rm Bind}$ [MeV] ~& $K_0$ [MeV] ~& $S_0$ [MeV] ~& $L_{0}$ [MeV]~\\
	\hline
	0.16 & 16 & 240 & 31 & 57.7\\
	\hline\hline
	\end{tabular}
	\label{saturation}
	\vspace{0.3cm}
\end{table}
We first use the masses of $\omega$ and $\rho$ mesons to fix $m_\omega$ and $m_\rho$ in  Eq.~(\ref{vector lag}).
The parameters $ cm_{u} = cm_d$ and $cm_{s}$ are fixed from $m_\pi f_\pi$ and $m_K f_K$ 
as 
\begin{align}
2cm_{u}=m_{\pi}^{2}f_{\pi}^{2}\,,\quad ~~~~
c ( m_{u} + m_{s} ) =m_{K}^{2}f_{K}^{2} \,.
\end{align}
The values of $g_1$ and $g_2$ are determined from the masses of nucleons at vacuum through Eq.~(\ref{N masses}) with $\sigma$ replaced with $f_\pi$.
There are still 9 parameters to be determined:
\begin{equation}
  \bar{\mu}^{2},\ \lambda_{4},\ \lambda_{6},\ \lambda_{8},\  \lambda_{10},\ B,\ \lambda_{\omega \rho} \,, \ g_{\omega NN},\ g_{\rho NN} \,.
\end{equation}
These parameters are tuned to reproduce the saturation properties listed in Table.~\ref{saturation}.
It turns out that there are some degeneracy related to the choice of parameters $\lambda_8, \lambda_{10}$, and $B$.
\begin{figure}[tbp]
\begin{center}
\includegraphics[width=6cm]{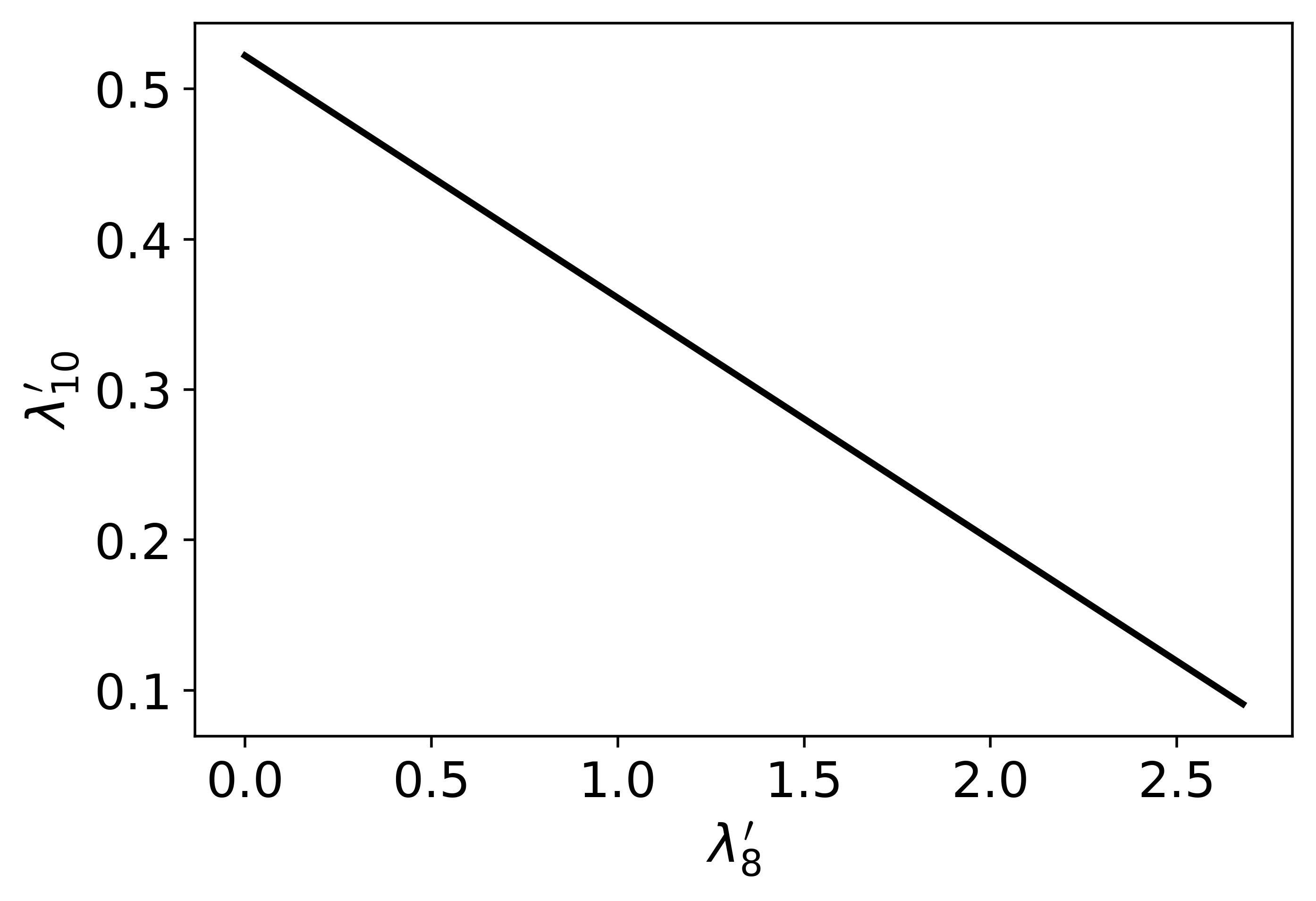}
\caption{Restricted combination of $\lambda_{8}$ and $\lambda_{10}$ after fixing the value of  $\sigma_{s}$ with $m_{0}=700\,\rm{MeV}$.  
We normalize the couplings as
$\lambda_{8}^{\prime}=\lambda_{8}f_{\pi}^{4}$ and $\lambda_{10}^{\prime}=\lambda_{10}f_{\pi}^{6}$.
}
\label{lamb8,lamb10}
\end{center}
\end{figure}
In Fig.\ref{lamb8,lamb10}, we show 
the range of $\lambda_8$ and $\lambda_{10}$ needed to satisfy the saturation properties.
Finally, we fit the parameter $B$ to reproduce the masses of $\eta$ and $\eta'$ mesons.
Here we omit the detail and show the determined values of model parameters only for $m_0=700\,$MeV as a typical example in Table~\ref{tab:my_label}.
We refer Ref.~\cite{Gao:2022klm} for the details of the determination and the values of model parameters for other choices of $m_0$.
%
\begin{table} [tb]
    \centering
\caption{  Model parameters determined from the saturation properties.     
When $B=600$ MeV, solutions satisfying
     the saturation properties can be found only  in the range: $0\leq\lambda_{8}^{'}\leq2.677$.
      Here we list the boundary values as typical examples;
    $\lambda_{8}'=0$ is the minimum boundary and $\lambda_{8}'=2.677$ is the maximum boundary. 
}
\begin{tabular}{c|c|c|c}
\hline \hline
&~ $m_{0}=700$ [MeV] ~&~ $\lambda_{8}'=0$ ~&~ $\lambda_{8}'=2.677$~  \\
\hline
 &$g_{1}$ & $7.81$ &  $7.81$ \\
 &$g_{2}$ & $14.26$  & $14.26$ \\
 &$\bar{\mu}^{2} / f_{\pi}^{2}$ & $23.21$  & $41.35$ \\
 &$\lambda_{4}$ & $133.4$  & $194.7$  \\
$B=600$ [MeV] ~& $\lambda_{6} f_{\pi}^{2}$& $82.71$   & $160.1$ \\
&$\lambda_{\omega \rho}$ & $0.3047$  & $0.3636$\\
&$\lambda_{10}f_{\pi}^{6}$ & $0.5221$  & $0.09091$ \\
&$g_{\omega N N}$ & $5.437$  & $5.142$ \\
&$g_{\rho N N}$ & $9.577$  & $9.541$ \\
\hline \hline
&$g_{1}$ & $7.81$ &  $7.81$ \\
   &$g_{2}$ & $14.26$  & $14.26$ \\
 &$\bar{\mu}^{2} / f_{\pi}^{2}$ & $39.98$  & $55.24$ \\
 &$\lambda_{4}$ & $94.02$  & $149.3$  \\
$B=0$ [MeV] ~& $\lambda_{6} f_{\pi}^{2}$& $62.23$   & $136.4$ \\
&$\lambda_{\omega \rho}$ & $0.2442$  & $0.2988$\\
&$\lambda_{10}f_{\pi}^{6}$ & $0.5221$  & $0.09091$ \\
&$g_{\omega N N}$ & $6.287$  & $5.957$ \\
&$g_{\rho N N}$ & $10.19$  & $10.21$ \\
\hline \hline
\end{tabular}
    \label{tab:my_label}
\end{table}


\begin{figure}[thb]
\centering
\includegraphics[width=6.5cm]{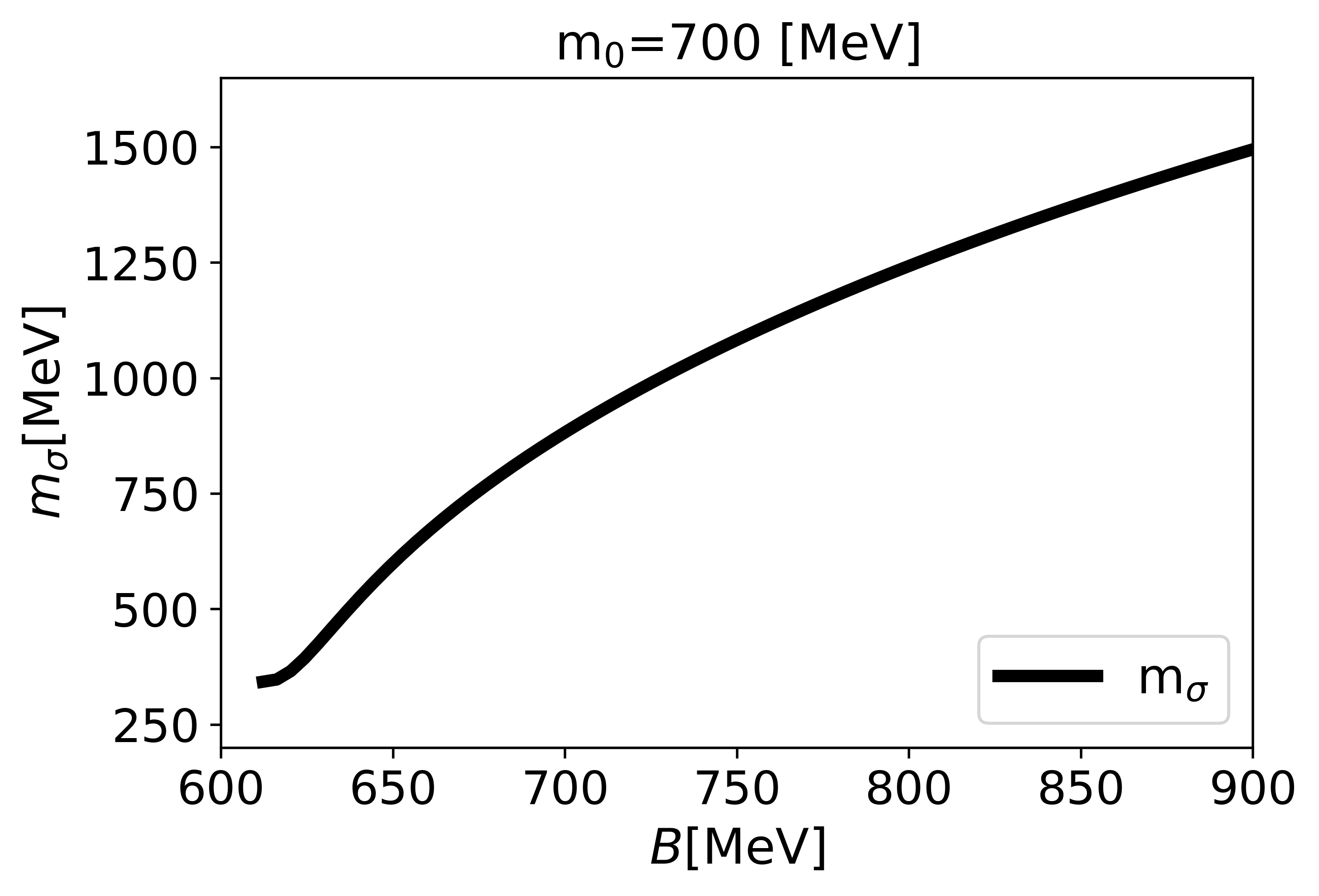}
\includegraphics[width=6.5cm]{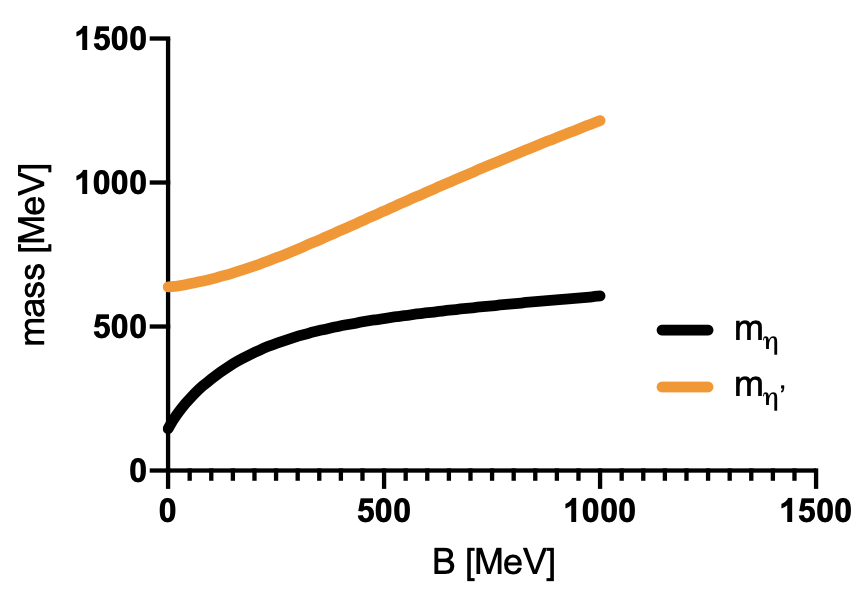}
\caption{
$B$ dependence of  $m_{\sigma}$ (left panel), $m_\eta$ and $m_{\eta'}$ (right panel) for $m_{0}=700 \,\rm{MeV}$.  
}
\label{vacuum sigma_mass}
\end{figure}

\subsection{Softening of the EOS by the Effect of Anomaly}
\label{sec:softening anomaly}

Here, we briefly explain the mechanism for the effect of anomaly to soften the EOS in the PDM. 
We refer Ref.~\cite{Gao:2022klm} for the details.

One of the key features is that both the condensate $\sigma$ and $\sigma_s$ are enhanced when the effect of anomaly is included. 
Their values at vacuum are actually increased with increasing $B$.
Since the mass of $\sigma$ meson, $m_\sigma$ is proportional to $\sigma$, the mass is increased as shown in Fig.\ref{vacuum sigma_mass}.
In the potential picture for nucleons, 
the $\sigma$ meson mediates the attractive force among nucleons in matter, so that
the larger $m_\sigma$ leads to shorter effective range of the attraction with the weaker overall strength.
The repulsive $\omega$ interaction should be weaker to balance the weaker attraction.
As a result, the weaker repulsion for a larger $B$ makes the EOS softer in the density region higher than the normal nuclear density.


\section{Quark matter from an NJL-type model}
\label{sec:QM NJL}

Following Ref.~\cite{Baym:2019iky}, 
we construct quark matter from 
an NJL-type effective model of quarks with the 4-Fermi interactions
which cause the color-superconductivity as well as  the spontaneous chiral symmetry breaking.
The Lagrangian is given by
\begin{align}\label{L_CSC}
	\lag_\CSC
	&=\lag_0
	+\lag_\sigma
	+\lag_\mathrm{d}
	+\lag_\mathrm{KMT}
	+\lag_\mathrm{vec} \ ,
\end{align}
where
\begin{align}
	\lag_0&=\bar q(i\gamma^\mu\partial_\mu-\hat m_q+\gamma_\mu \hat{A}^\mu)q \ , \\
	\lag_\sigma&=G\sum_{A=0}^8\qty[(\bar q\tau_Aq)^2+(\bar qi\gamma_5\tau_Aq)^2]\ , \\
	\lag_\mathrm{d}
	&=H\sum_{A,B=2,5,7}\left[(\bar q\tau_A\lambda_BC\bar q^t)(q^tC\tau_A\lambda_Bq)\right. 
+\left.(\bar qi\gamma_5\tau_A\lambda_BC\bar q^t)(q^tCi\gamma_5\tau_A\lambda_Bq)\right]\ , \\
	\lag_\mathrm{KMT}
	&=-K\qty[\det_f\bar q(1-\gamma_5)q+\det_f\bar q(1+\gamma_5)q]\ , \label{KMT term} \\
	\lag_\mathrm{vec}
	&=-g_V(\bar q\gamma^\mu q)(\bar q\gamma_\mu q) \ ,
\end{align}
with $\hat{A}^\mu$ being the collection of chemical potentials 
\begin{align}
	\hat{A}^\mu = (\mu_q+\mu_3\lambda_3+\mu_8\lambda_8+\mu_QQ)\delta^\mu_0 \ \,.
\end{align}
Here $\lambda_a$ are Gell-Mann matrices in color space,
while $\tau_0= {\bf 1}_{3\times 3} \sqrt{2/3} $ and $\tau_A (A=1\cdots8)$ are the Gell-Mann matrices 
and $Q=\frac{1}{2}\tau_3 + \frac{1}{2\sqrt{3}}\tau_8 = \diag{2/3,-1/3,-1/3}$ 
is a charge matrix in flavor space. 
Meanwhile $\tau_0= {\bf 1}_{3\times 3} \sqrt{2/3} $ and $\tau_A (A=1\cdots8)$ are the Gell-Mann matrices for the flavor.
For coupling constants $G$ and $K$ as well as the cutoff $\Lambda$, we use
$G\Lambda^2=1.835$, $K\Lambda^5=9.29$ and $\Lambda=631.4$\,MeV, 
which successfully reproduce the hadron phenomenology at low energy~\cite{Hatsuda:1994pi,Baym:2017whm}.
The mean fields are introduced as
\begin{align}
\sigma_f & = \ev {\bar{q}_f q_f} \ , \quad ( f =u, d, s ) \ , \\
d_j & = \ev {q^tC\gamma_5R_jq} \ , \quad ( j = 1,2,3 ) \ , \\
n_q & = \sum_{f=u,d,s} \ev{ q_f^\dag q_f } \ ,
\end{align}
where 
$(R_1,R_2,R_3)=(\tau_7\lambda_7,\tau_5\lambda_5,\tau_2\lambda_2)$. 
The resultant thermodynamic potential is calculated as
\begin{align}
	\Omega_\CSC 
	&=\Omega_s-\Omega_s[\sigma_f=\sigma^0_f,d_j=0,\mu_q=0] 
+\Omega_c-\Omega_c[\sigma_f=\sigma^0_f,d_j=0]\ , 
\end{align}
where
\begin{align}
	\Omega_s&=-2\sum_{\alpha=1}^{18}\int^\Lambda_{\mathbf{p}}\frac{\varepsilon_\alpha}{2}\ , \label{Omega s}\\
	\Omega_c&=\sum_{f=u,d,s}2G\sigma_i^2+\sum_{j=1,2,3} Hd_j^2-4K\sigma_u\sigma_d\sigma_s-g_Vn_q^2\ .
\end{align}
In Eq.~(\ref{Omega s}), 
$\varepsilon_\alpha$ are energy eigenvalues of the 
inverse propagator in Nambu-Gor'kov basis given by 
\begin{align}
	S^{-1}(k)&=\mqty(\gamma_\mu k^\mu-\hat M+\gamma^0\hat\mu & \gamma_5\sum_i\Delta_iR_i \\
	-\gamma_5\sum_i\Delta_i^\ast R_i & \gamma_\mu k^\mu-\hat M-\gamma^0\hat\mu)\ , 
\label{inverse propagator}
\end{align}
where
\begin{align}
	M_i&=m_i-4G\sigma_i+K \sum_{j,k=u,d,s}  |\epsilon_{ijk}|\sigma_j\sigma_k\ , \quad (i = u,d,s) \ ,  \\
	\Delta_j&=-2Hd_j\ , (j=1,2,3) \ ,\\
	\hat\mu&=\mu_q-2g_Vn_q+\mu_3\lambda_3+\mu_8\lambda_8+\mu_QQ \ .
\end{align}
The inverse propagator 
$S^{-1}(k)$ in Eq.~(\ref{inverse propagator}) is $72\times72$ matrix 
in terms of the color, flavor, spin and Nambu-Gorkov basis and has 72 eigenvalues.
$M_{u,d,s}$ are the constituent masses of the $u,d,s$-quarks 
and $\Delta_{1,2,3}$ are the color-superconducting 
gap energies. 
In the high density region, $n_B \gtrsim 5 n_0$, their ranges are 
$M_{u,d}\approx50$--$100$ MeV, $M_s\approx$ 250--300 MeV 
and $\Delta_{1,2,3}\approx$ 200--250 MeV~\cite{Baym:2017whm}. 
We note that the inverse propagator matrix does not depend on the spin, and that the charge conjugation invariance relates two eigenvalues.
Then, there are 18 independent eigenvalues at most.

The entire thermodynamic potential is constructed by adding the lepton contribution in Eq.~(\ref{pot: lepton}) as 
\begin{align}
	\Omega_\quark=\Omega_\CSC+\sum_{l=e,\mu}\Omega_l \ .
\end{align}
The chiral condensates $\sigma_i$ ($i=u,d,s$) and the diquark condensates $d_j$ ($j=1,2,3$) are determined from the gap equations:
\begin{align}
	\pdv{\Omega_\quark}{\sigma_i}=0 \ , \quad \pdv{\Omega_\quark}{d_j} = 0  \ .
\end{align}
The relevant chemical potentials other than the baryon number density are determined from the beta equilibrium condition 
in Eq.~(\ref{beta eq.})  combined with 
the conditions for electromagnetic charge neutrality and color charge neutrality expressed as
\begin{align}
	n_j=-\pdv{\Omega_\quark}{\mu_j}=0 \ , \quad (j = 3, 8, Q) \ .
\end{align}
The baryon number density $n_B$ is equal to  three times of quark number density given by 
\begin{align}
	n_q=-\pdv{\Omega_\quark}{\mu_q} \ , 
\end{align}
where $\mu_q$ is the quark number chemical potential 
which is $1/3$ of the baryon number chemical potential.
Substituting the above conditions, we obtain the pressure of the system as
\begin{equation}\label{P_Q}
	P_\quark=-\Omega_\quark \ .
\end{equation}

\subsection{Softening of the EOS by the Effect of Anomaly in NJL-type model}
\label{sec:softening anomaly in NJL}

\begin{figure}[thb]
\centering
\includegraphics[width=6.5cm]{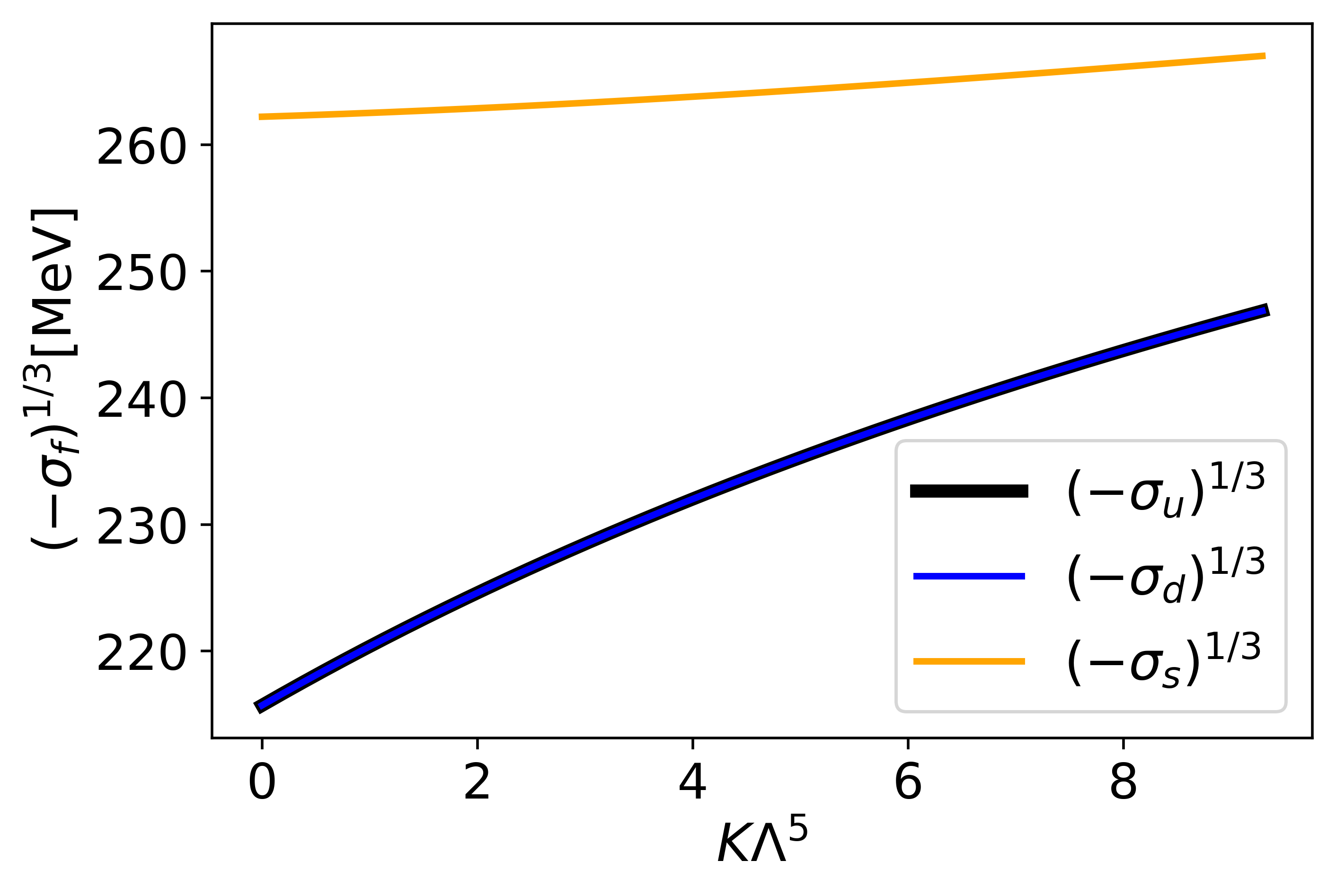}
\includegraphics[width=6.5cm]{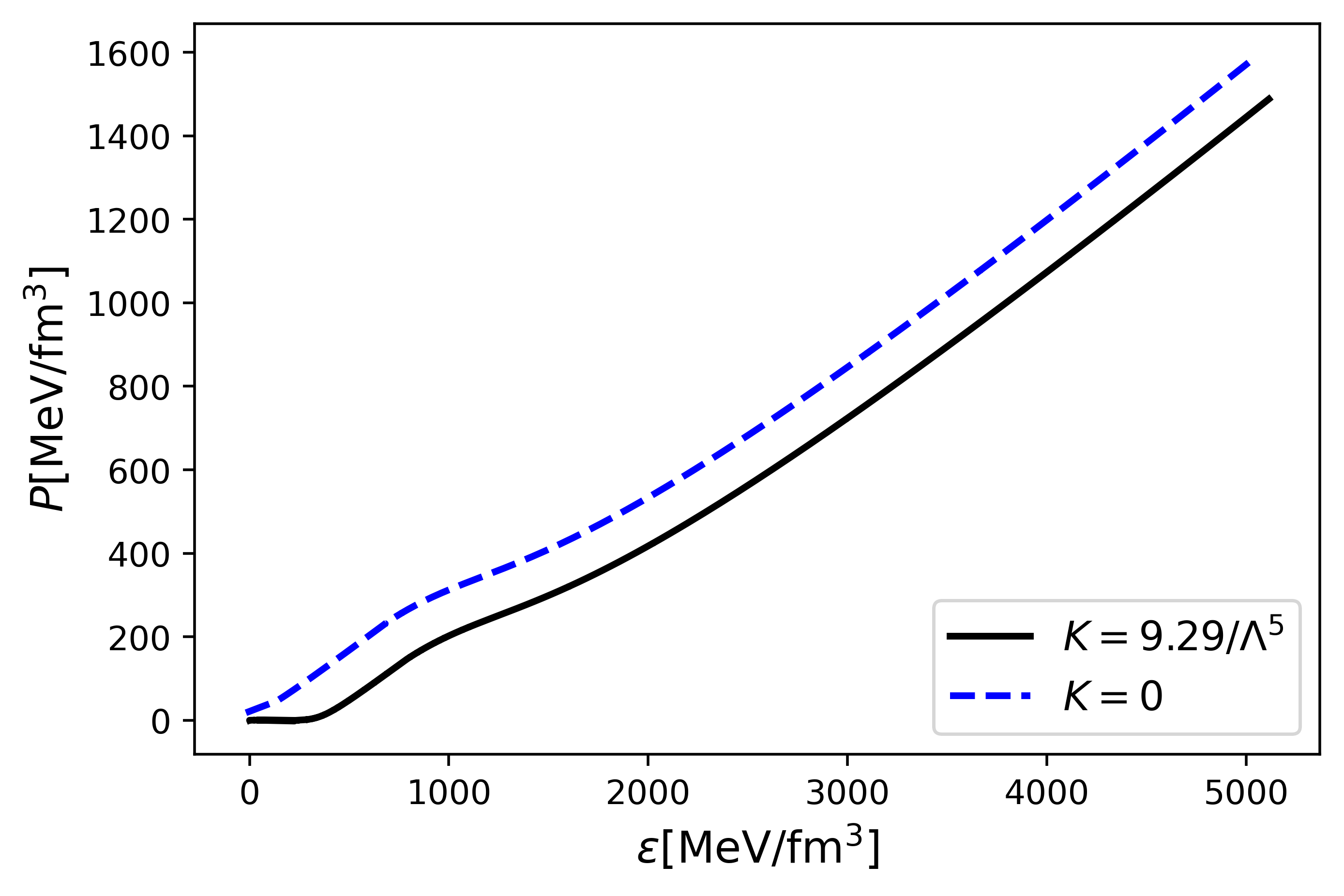}
\caption{
$K$ dependence of chiral condensates (left panel) and the energy dependence of pressure for $H/G = 0$ and $g_{V}/ G =0.1$ (right panel).
}
\label{Anomaly_NJL}
\end{figure}

Here we briefly explain how the anomaly softens the EOS in the NJL-type quark model.
For simplicity we set $H=0$ and omit the effects of diquarks.
In the KMT interaction in Eq.~(\ref{KMT term}), 
the coefficient $K$ represents the strength of U(1)$_A$ anomaly. 
The anomaly assists the chiral symmetry breaking and lowers the ground state energy in vacuum;
a larger $K$ leads to chiral condensates greater in magnitude as shown in the left panel of Fig.\ref{Anomaly_NJL}. 

With the chiral restoration, the system loses the energetic benefit of having the chiral condensates.
Such release of the energy is more radical with the anomaly than without it.
As one can see from the thermodynamic relation $P = -\varepsilon + \mu_{q}n_{q}$,
the larger energy $\varepsilon$ with a stronger anomaly leads to the smaller pressure, i.e., the softening.
In other words, with the anomaly we have to add a larger ``bag constant'' to the energy density but must subtract it from the pressure.
We show  the resulting EOSs for $H/G = 0$ and $g_{V}/ G =0.1$ in the right panel of Fig.~\ref{Anomaly_NJL}.


\section{Interpolated EOSs and $M$-$R$ relations of NSs}
\label{sec:interpolation}

\subsection{Interpolation of EOSs}

In this subsections, we briefly explain how to interpolate the EOS for hadronic matter to that for quark matter constructed in previous sections.
Following Ref.~\cite{Baym:2017whm}, 
we assume that hadronic matter is realized
in the low density region
$n_B<2n_0$, 
and use the pressure constructed in Eq.~(\ref{P_H}).
In the high density region $n_B>5n_0$, 
the pressure given in Eq.~(\ref{P_Q}) of quark matter is used. 
In the intermediate region 
$2n_0<n_B<5n_0$, 
we assume that the pressure is expressed by a fifth order polynomial of $\mu_B$ 
as\footnote{
It is important to make interpolation for a correct set of variables, either $P(\mu_B)$ or $\varepsilon(n_B)$, from which one can deduce all the thermodynamic quantities by taking derivatives \cite{Baym:2017whm}. 
Other combinations, e.g., $P(\varepsilon)$ can not be used to derive $n_B$ and hence would miss some constraints.
}
\begin{align}
	P_\interp(\mu_B)=\sum_{i=0}^5C_i\mu_B^i \ .
\end{align}
Following the quark-hadron continuity scenario, we demand the interpolating EOS to match quark and hadronic EOS up to the second derivatives (otherwise we would have the first or second order phase transitions at the boundaries).
The six parameters $C_i$ ($i=1,\ldots,6$) are determined from the boundary conditions given by
\begin{align}
\frac{ \dd^n P_{ {\rm I}} }{ (\dd \mu_B )^n } \bigg|_{ \mu_{ BL} } =\frac{ \dd^n P_{\hadron} }{ (\dd \mu_B)^n } \bigg|_{ \mu_{ BL} } \,, \quad 
\frac{ \dd^n P_{ {\rm I}} }{ (\dd \mu_B )^n } \bigg|_{ \mu_{ BU} } =\frac{ \dd^n P_{\quark} }{ (\dd \mu_B)^n } \bigg|_{ \mu_{ BU} } \,,\quad (n=0,1,2) \,, 
\end{align}
where $\mu_{BL}$ is the chemical potential corresponding to $n_B=2n_0$ and $\mu_{BU}$ to  $n_B=5n_0$.

In addition to these boundary conditions,
the interpolated pressure must obey the causality constraint, i.e., the sound velocity, 
\begin{align}
	c_s^2=\dv{P}{\varepsilon}=\frac{n_B}{\mu_B\chi_B} \ ,
\end{align}
where 
$n_B=\dv{P}{\mu_B}$ and 
$\chi_B=\dv[2]{P}{\mu_B}$,
to be less than the light velocity.
This condition is more difficult to be satisfied
for the combination of softer nuclear EOS and stiffer quark matter EOS,
since such soft-to-stiff combination requires a larger slope in $P(\varepsilon)$.

We show an example of the interpolated pressure in Fig.~\ref{P-muB} 
with a parameter set
$\lambda'_8 = 2.677, \lambda'_{10}=0.09091$ for $m_0=700\,$MeV and $B=600\,\mbox{MeV}$ for the PDM,
and two parameter sets $(H/G, g_V/G)=(1.45, 0.4)$ and $(1.45, 0.5)$ for quark matter. 
Both plots \ref{P: 700-sat} and \ref{P: 700-unsat} in Fig.~\ref{P-muB} are smoothly connected by construction, 
but the set $(H/G, g_V/G)=(1.45, 0.4)$ violates causality as seen in Fig.~\ref{cs2-muB}, 
and therefore must be excluded. 

The $c_s^2$ exceeding the conformal value, $c_s^2=1/3$, and subsequent reduction within the interval $2$-$5n_0$ 
is the characteristic feature of the crossover models \cite{Masuda:2012kf,Masuda:2012ed,Baym:2017whm,Baym:2019iky,Kojo:2021wax}.
In nuclear domain, the sound velocity is small, $c_s^2 \sim 0.1$, while the natural size is $c_s^2 \sim 1/3$ in quark matter. 
In the intermediate region $c_s^2$ makes a peak.
How to approach the conformal limit is the subject under intensive discussions, 
see Refs.~\cite{Rho:2022wco,Fujimoto:2022ohj,Marczenko:2022jhl,Ivanytskyi:2022bjc,Pisarski:2021aoz}.

%
\newcommand{\figsize}{0.49\hsize}
\newcommand{\widthsize}{\hsize}
\begin{figure}\centering
	\begin{subfigure}{\figsize}
		\includegraphics[width=\widthsize]{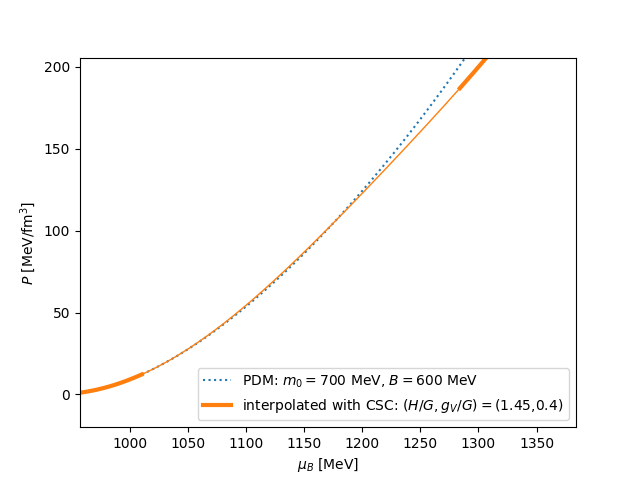}
		\caption{$(H/G,g_V/G)=(1.45,0.4)$}
		\label{P: 700-sat}
	\end{subfigure}
	\begin{subfigure}{\figsize}
		\includegraphics[width=\widthsize]{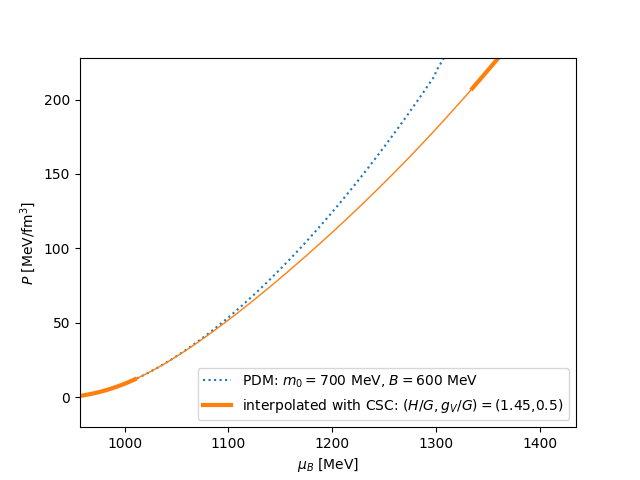}
		\caption{$(H/G,g_V/G)=(1.45,0.5)$}
		\label{P: 700-unsat}
	\end{subfigure}
\caption[]{
Pressure $P(\mu_B)$ of the PDM and the unified equations of state.
For the PDM we chose $\lambda'_8 = 2.677, \lambda'_{10}=0.09091$ for $m_0=700\,$MeV and $B=600\,\mbox{MeV}$  as a typical parameter set 
and for quark models we used $(H/G,g_V/G)=(1.45,0.4)$ and $(1.45,0.5)$. 
The thick curves in the unified equations of state are used to mark the pure hadronic and quark parts.  
}
\label{P-muB}
\end{figure}
%
\begin{figure}\centering
	\begin{subfigure}{\figsize}
		\includegraphics[width=\widthsize]{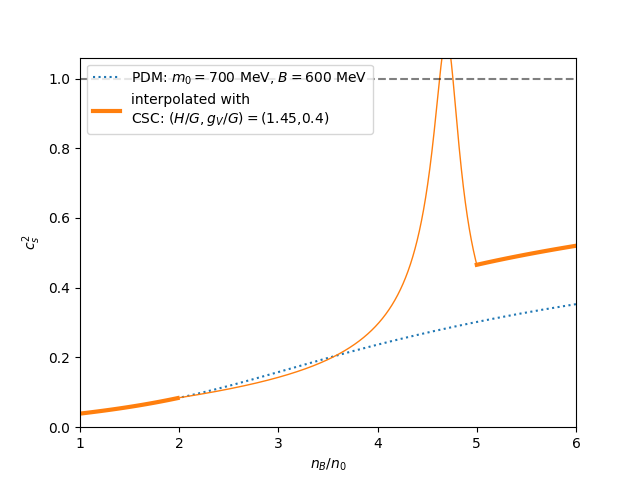}
		\caption{$(H/G,g_V/G)=(1.45,0.4)$}
		\label{cs: 700-sat}
	\end{subfigure}
	\begin{subfigure}{\figsize}
		\includegraphics[width=\widthsize]{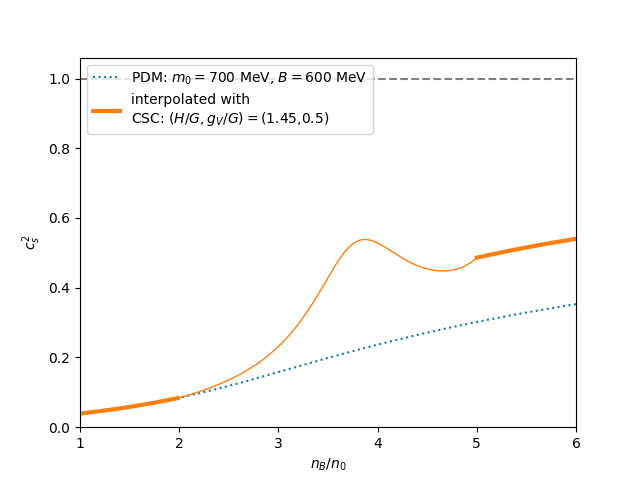}
		\caption{$(H/G,g_V/G)=(1.45,0.5)$}
		\label{cs: 700-unsat}
	\end{subfigure}
\caption[]{
Squared speed of sound $c_s^2$ 
for $(H/G,g_V/G)=(1.45,0.4)$ and (1.45,0.5). 
Curves are same as in Fig.~\ref{P-muB}. 
}
\label{cs2-muB}
\end{figure}

\begin{figure}\centering
	\begin{subfigure}{\figsize}
		\includegraphics[width=\widthsize]{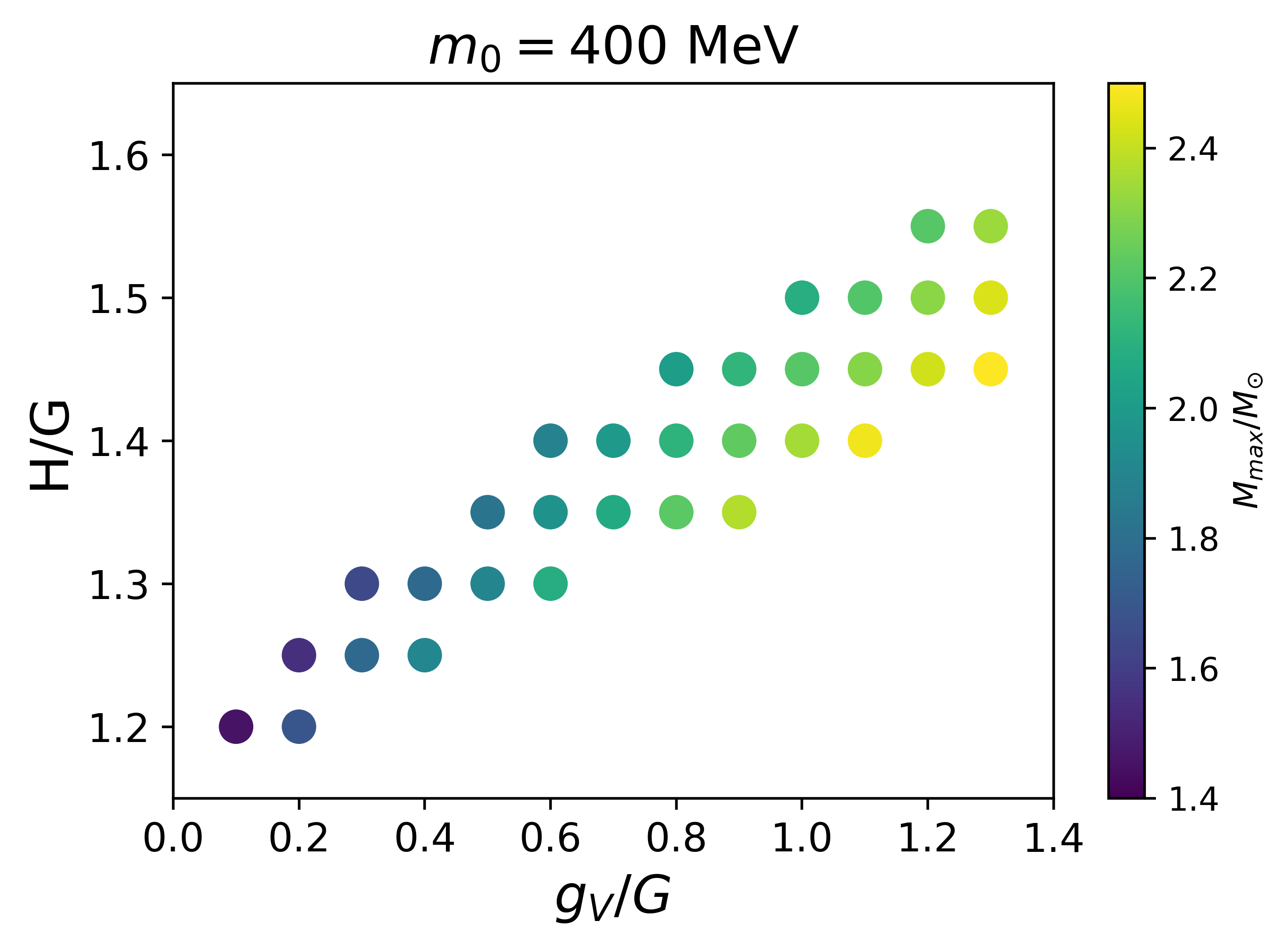}
		\caption{}
	\end{subfigure}
			\vspace{0.4cm}
	\begin{subfigure}{\figsize}
		\includegraphics[width=\widthsize]{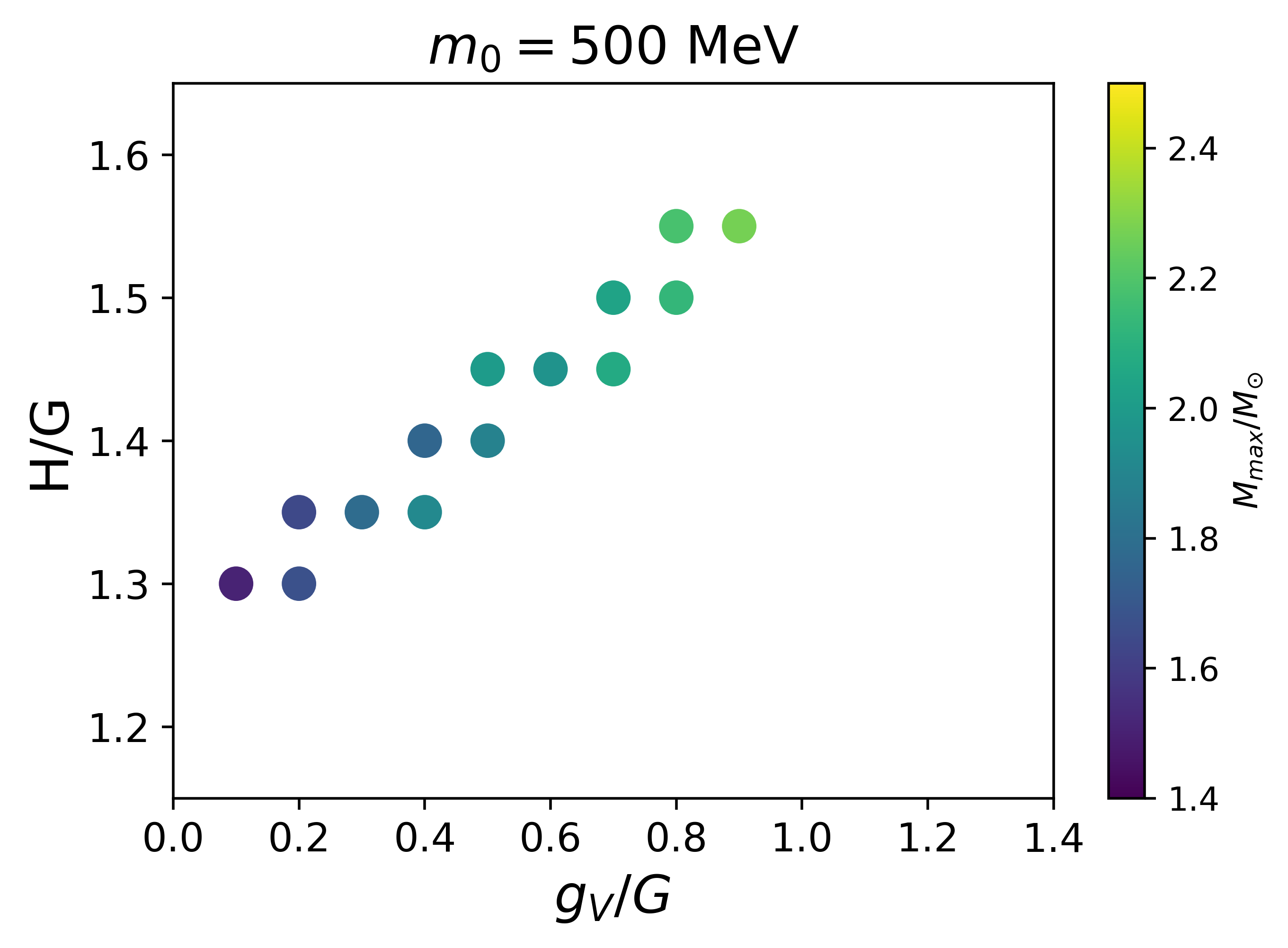}
		\caption{}
	\end{subfigure}
		\vspace{0.4cm}
	\begin{subfigure}{\figsize}
		\includegraphics[width=\widthsize]{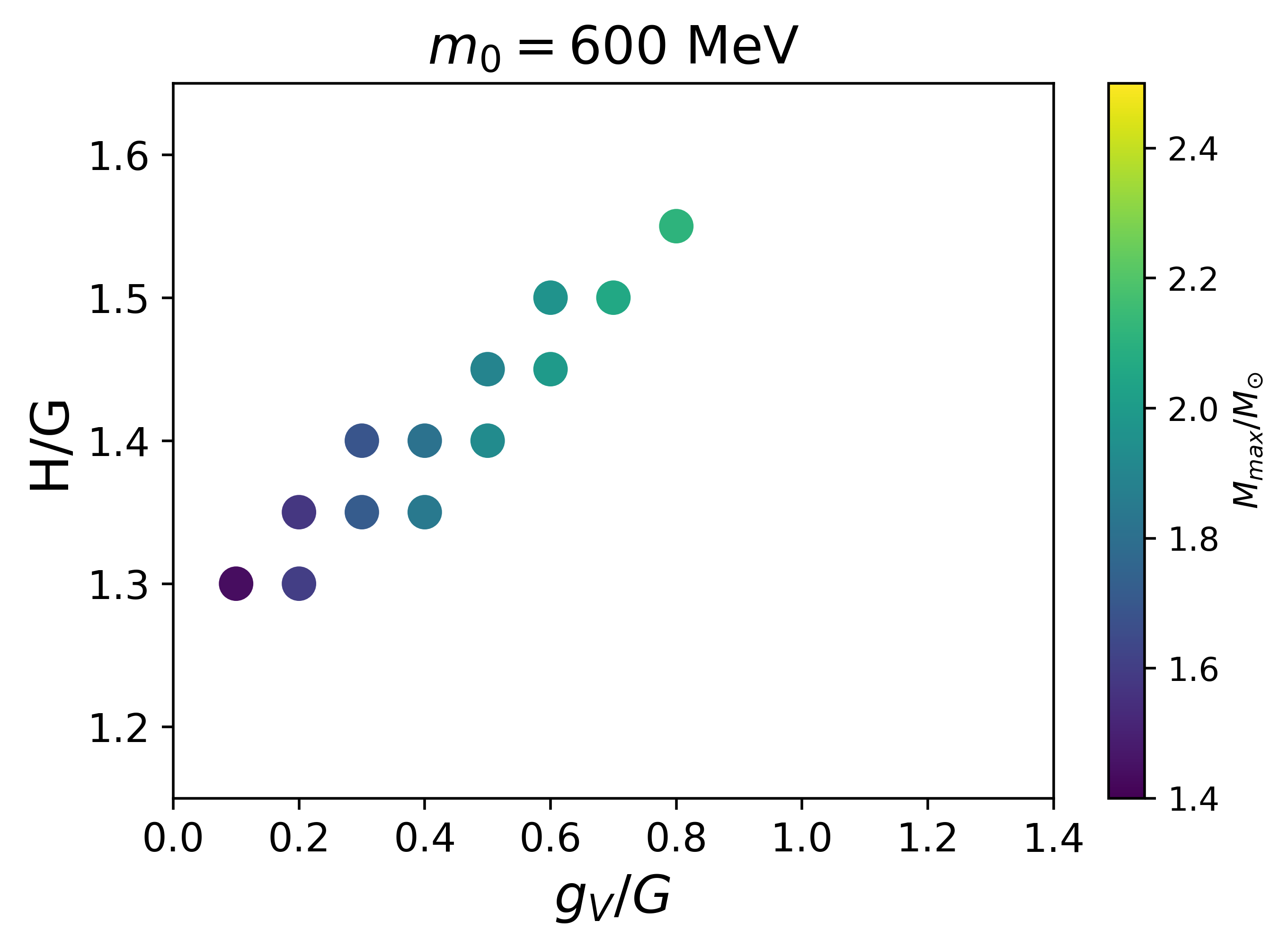}
		\caption{}
	\end{subfigure}
	\begin{subfigure}{\figsize}
		\includegraphics[width=\widthsize]{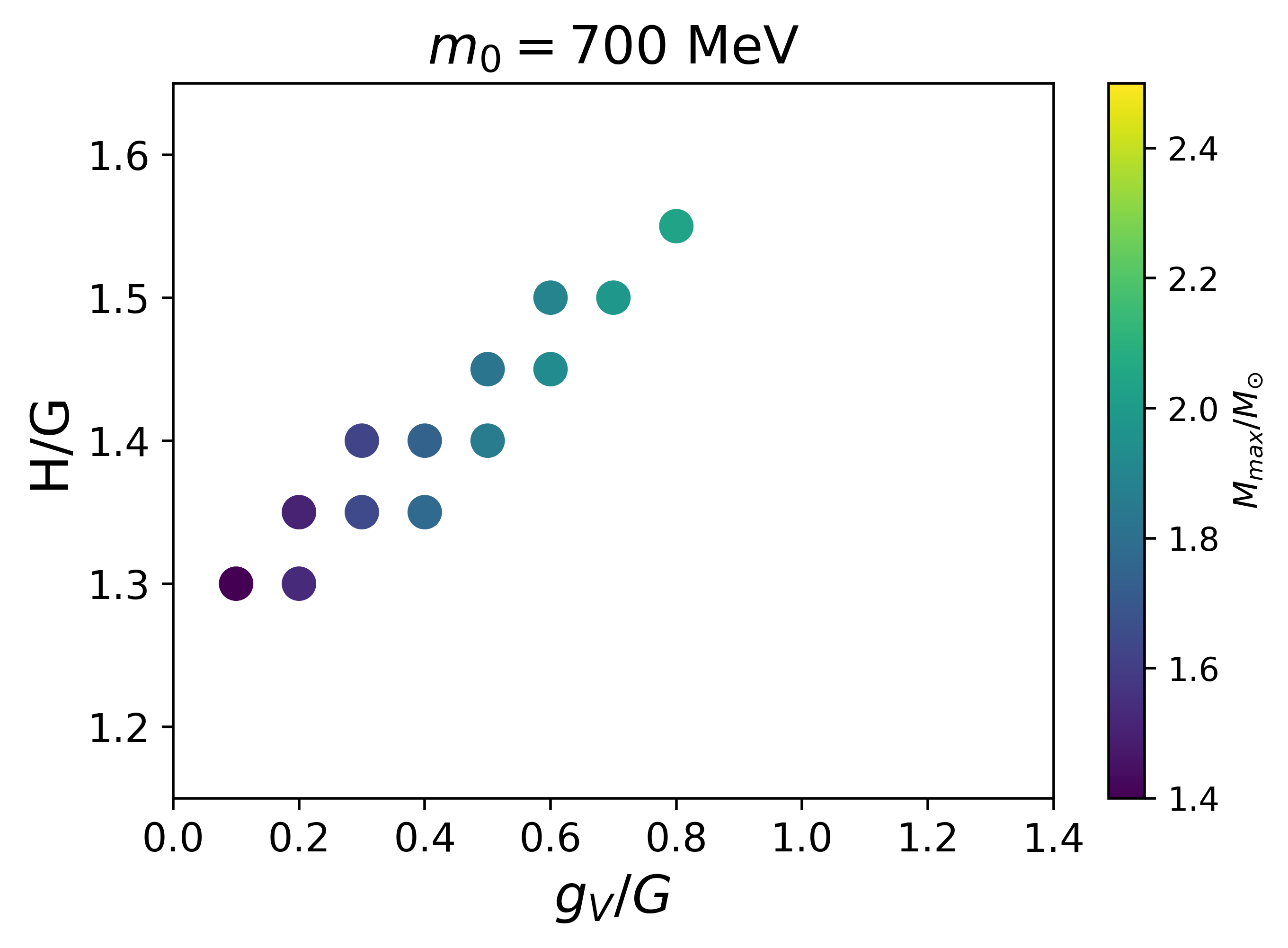}
		\caption{}
	\end{subfigure}
		\vspace{0.4cm}
	\begin{subfigure}{\figsize}
		\includegraphics[width=\widthsize]{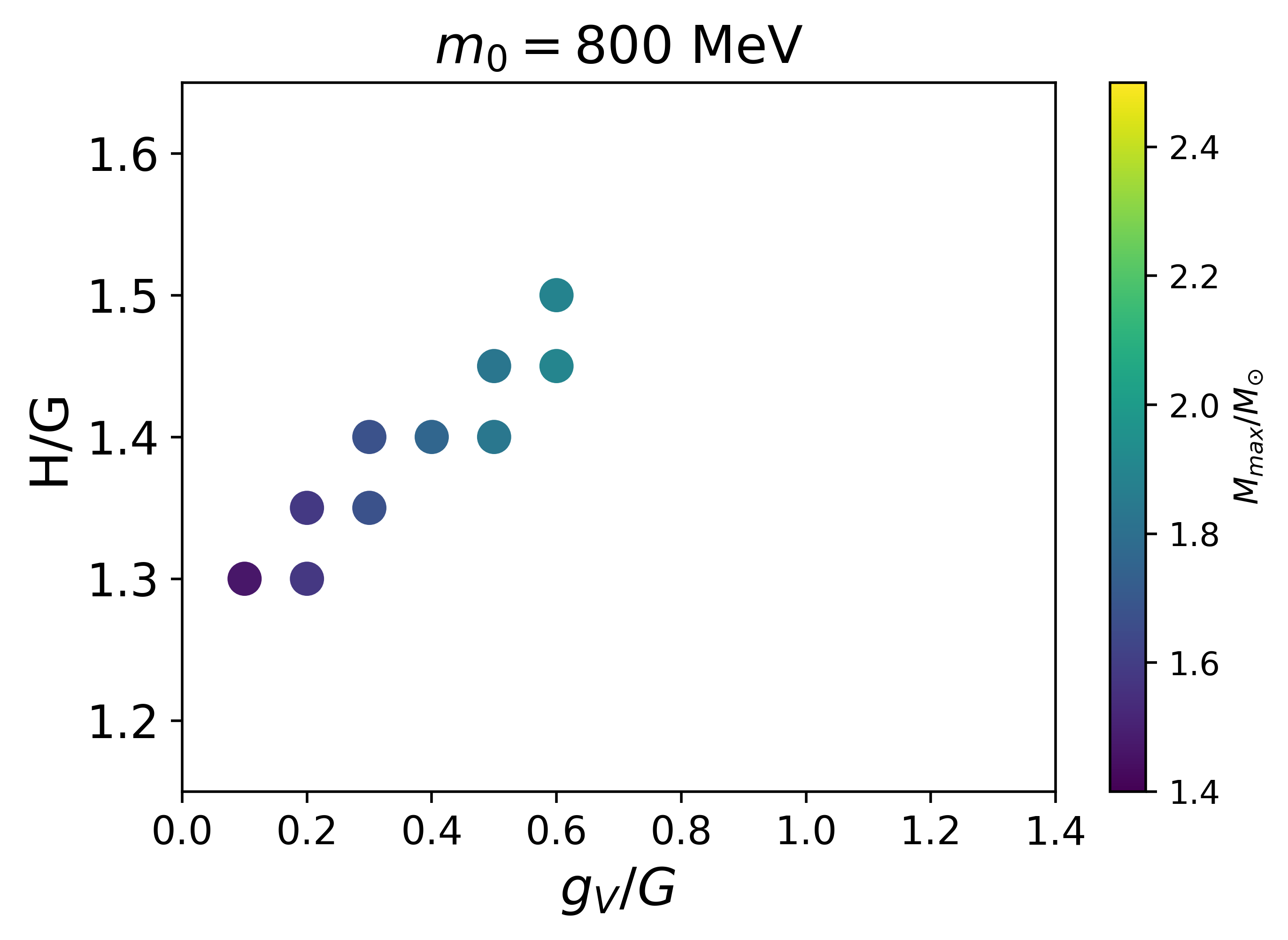}
		\caption{}
	\end{subfigure}
\caption{ 
Allowed combinations of $(H,g_V)$ for $m_0=400\text{--}800\,\mathrm{MeV}$. 
The color of the circle shows the maximum mass of neutron stars obtained
from the corresponding parameters, 
as indicated by a vertical bar at the right side of each figure. 
}
\label{cond_check_with_Mmax}
\end{figure}

Figure~\ref{cond_check_with_Mmax} shows 
allowed combinations of $(H,g_V)$ for several choices of $m_0$.  
%
Here, we fix the parameters in the PDM to $B=600$\,MeV and $\lambda_{8}'=0$, 
which determines the value of $\lambda_{10}'$ as summarized in sec.~\ref{sec:parameters} (e.g. $\lambda_{10}'=0.5221$ for $m_0=700$\,MeV).
The parameter $\lambda_{\omega\rho}$ is set to reproduce the slope parameter as $L_0=57.7$ MeV.
In all cases, the allowed values of $H$ and $g_V$ have a positive correlation;
for a larger $g_V$ we need to increase the value of $H$ \cite{Baym:2019iky}. For $m_{0}=800$ MeV, the maximum masses for all the combinations are below $2M_{\odot}$, leading to the conclusion that $m_{0}=800$ MeV should be excluded within the current setup of the PDM parameters.
The details of the positive correlation between $H$ and $g_{V}$ depend on the low density constraint and the choice of $m_0$.
As we mentioned in Introduction, the EOS in hadronic matter is softer for a larger $m_0$.
Correspondingly, the parameter $g_V$, which makes quark matter EOS stiff, should not be too large for causal interpolations;
for a larger $m_0$, the acceptable $g_V$ tends to appear at lower values.
Typical values of $(H,g_V)$ are greater than expected from the Fierz transformation for which $(H,g_V)=(0.5, 0.5)G$ (see e.g.~Ref. \cite{Buballa:2003qv}).
Such choices were used in the hybrid quark-hadron matter EOS with first order phase transitions, 
but they tend to lead to predictions incompatible with the $2M_\odot$ constraints.

\subsection{$M$-$R$ relations of NSs}

With the unified EOS explained so far,
we now calculate $M$-$R$ relations of NSs by solving 
the Tolman-Oppenheimer-Volkoff (TOV) equation~\cite{Tolman:1939jz,Oppenheimer:1939ne},
\begin{align}
	\begin{aligned}
	\dv{P}{r}&=-G\frac{(\varepsilon+P)(m+4\pi r^3P)}{r^2-2Gmr} \ , \\
	\dv{m}{r}&=4\pi r^2\varepsilon \ , 
	\end{aligned}
\end{align}
where $G$ is the Newton constant, 
$r$ is the distance from the center of a neutron star, 
$P$, $m$ and $\varepsilon$ are the pressure, mass, and energy density 
as functions of $r$:
\begin{align}
	P=P(r) \ , \quad m=m(r) \ , \quad \varepsilon=\varepsilon(r) \ .
\end{align}
The radius $R$ is determined by the condition $P(R)=0$ and the mass $M$ by $M=m(R)$.
To estimate radii of NSs in accuracy better than $\sim 0.5$ km, we need to include the crust EOS. 
We use the BPS EOS \cite{Baym:1971pw} for the outer and 
inner crust parts.\footnote{
The BPS EOS is usually referred as EOS for the outer crust, but it also contains the BPP EOS \cite{Baym:1971pw} for the inner crust. 
at $n_B \leq 0.1\,\mathrm{fm}^{-3}$,
and at $n_B \geq 0.1\,\mathrm{fm}^{-3}$ we use our unified EOS from nuclear liquid to quark matter.
For a given central density, we obtain the corresponding $M$-$R$ point, and the sequence of such points form the $M$-$R$ curves.
}

In order to study the relation between microscopic parameters and $M$-$R$ relations,
below we examine the impacts of the PDM EOS, the dependence on the $\omega^2 \rho^2$ coupling ($\lambda_{\omega\rho}$), 
the chiral invariant mass $m_0$, and the anomaly strength $B$ for a given set of quark matter parameters $(H,g_V)$. 

\begin{figure}
\centering
\includegraphics[width=8cm]{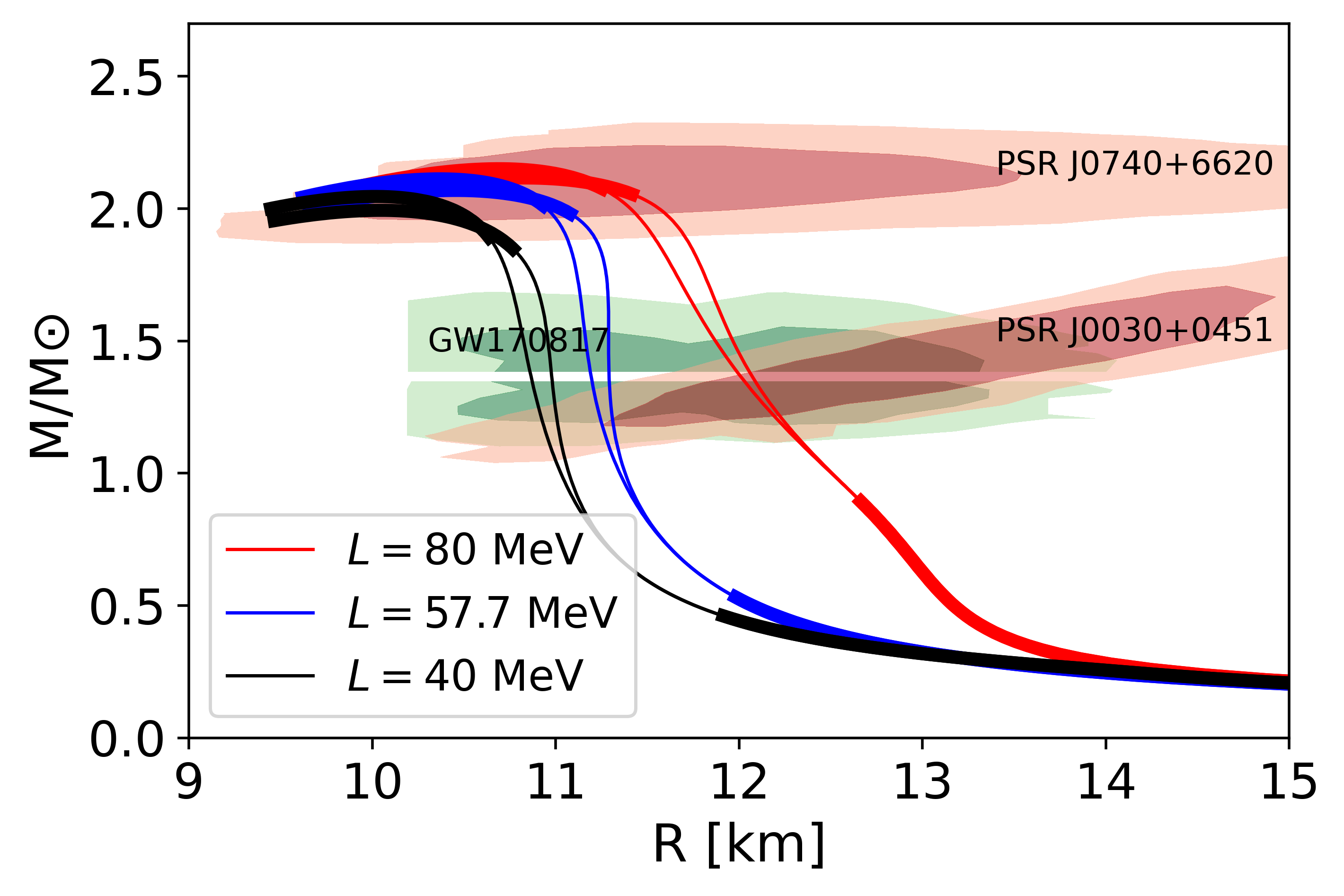}
\caption{Dependence of $M$-$R$ relations for  $m_{0}=500~$MeV on the slope parameter. 
Red curves are connected to the NJL parameters $(H, g_V)/G=$ (1.55, 1.0), (1.50, 0.9);
blue curves to (1.55, 0.9), (1.50, 0.8); black curves to (1.55, 0.8), (1.50, 0.7).
}
\label{gv_H}
\end{figure}

We first study the effect of the $\omega^{2}\rho^{2}$ interaction. 
We fix $m_0=500$\,MeV and $B = 600$\,MeV , and vary $\lambda_{\omega\rho}$ which leads to changes in the slope parameter $L_0$ in the symmetry energy.
We examine the cases with $L_0=40, 57.7$, and $80$ MeV since  the value of $L_0$ still has the uncertainty which is being intensively studied~\cite{Lattimer:2023rpe}. 
The resultant $M$-$R$ relation is shown in Fig.~\ref{gv_H}.
The $M$-$R$ relations with the core density smaller than $2n_0$ (larger than 5$n_0$) are emphasized by thick curves in the low (high) mass region.
The $\lambda_{\omega \rho} > 0 $ corresponds to attractive correlations that reduce $L_0$ and soften EOS in the nuclear domain.
For $L_0=40, 57.7$, and $80$\,MeV, the radii of $1.4 M_\odot$ NS are $\simeq 11.05$ km,
$\simeq 11.2$ km, and $\simeq 12.1$ km, respectively 
%
Precise determination of slope parameter in the future will help us to further constrain the NS properties, especially the radii.
%

\begin{figure}
		\includegraphics[scale=0.48]{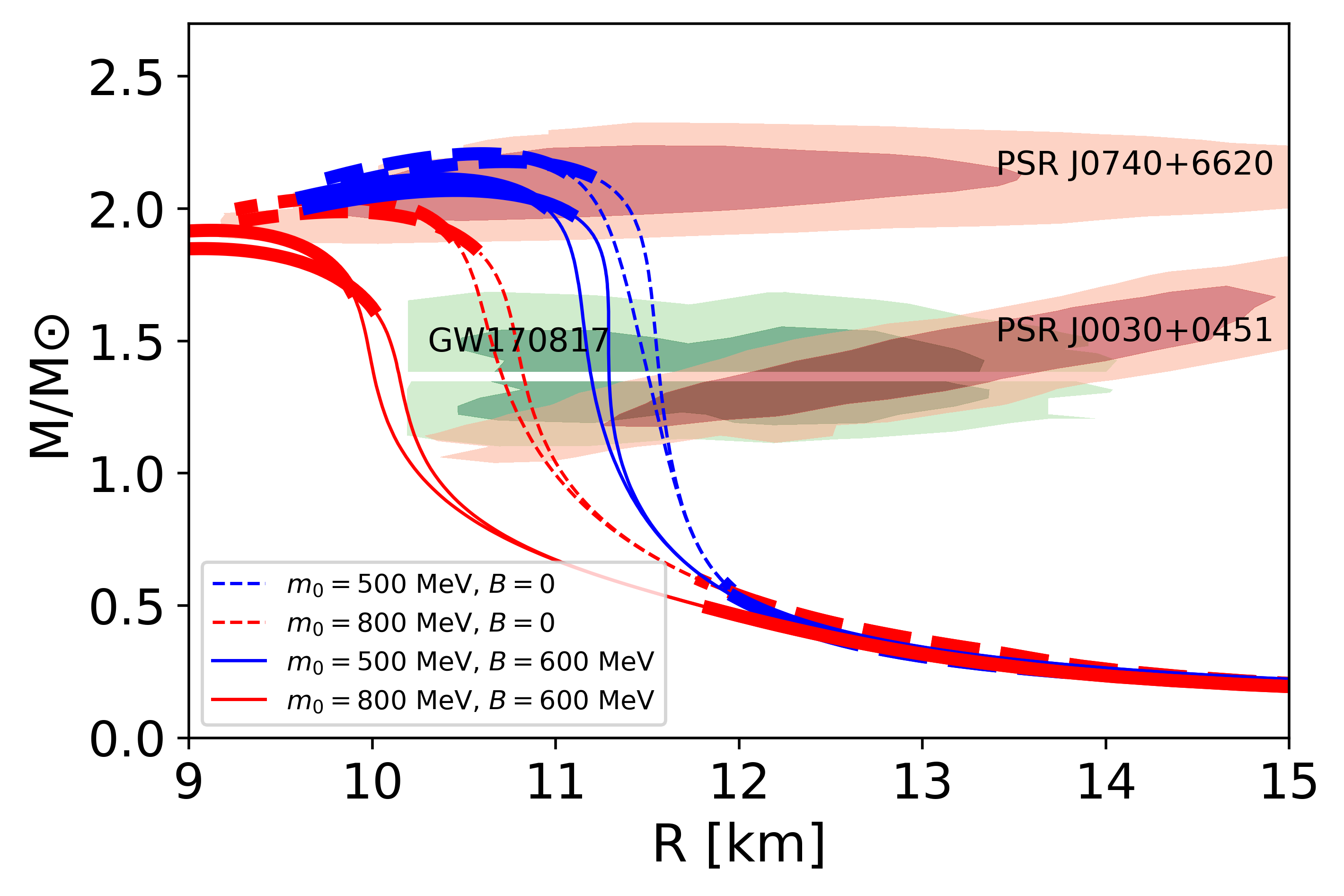}
\ 
		\includegraphics[scale=0.48]{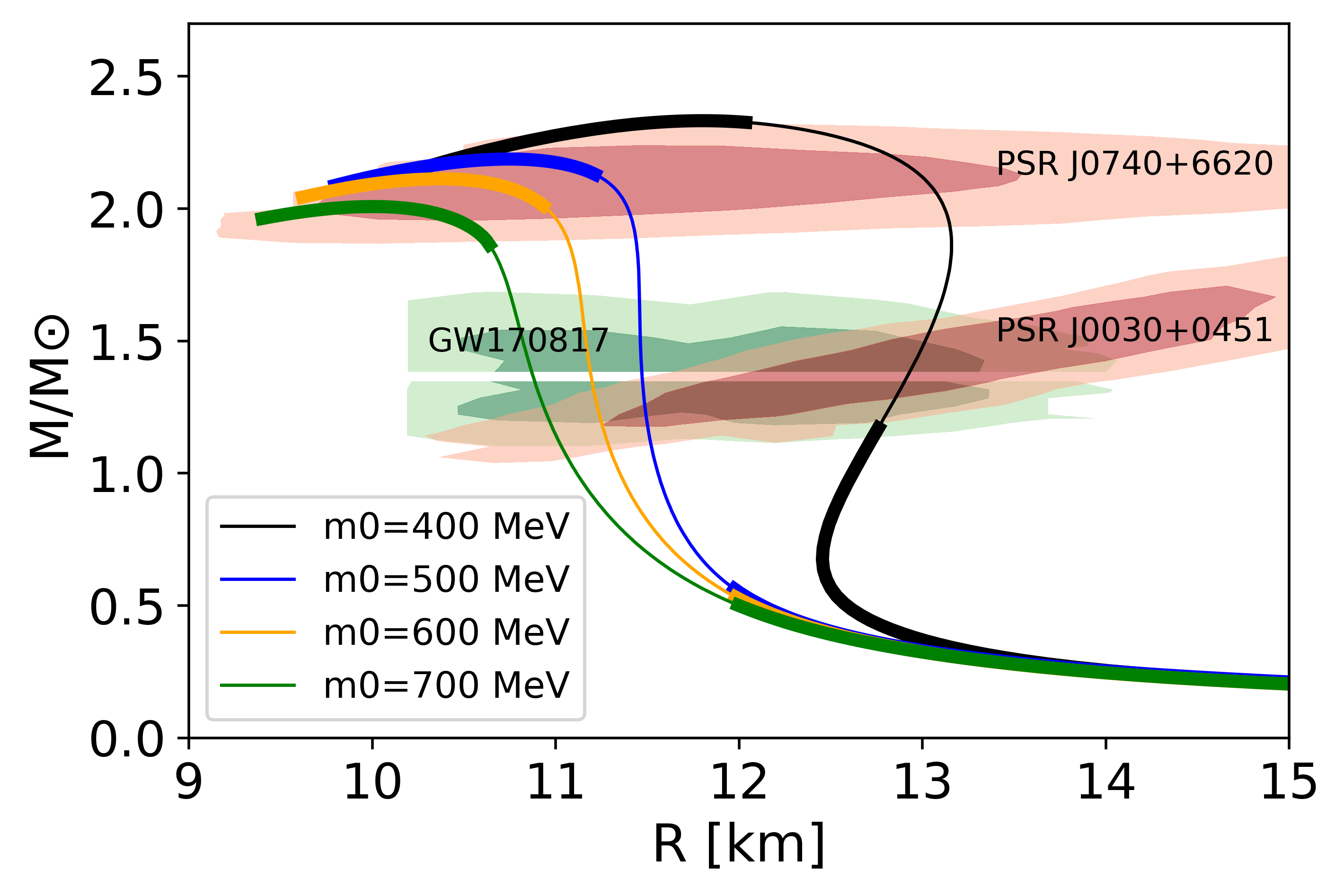}
\caption{
Mass-radius relations for different $m_{0}$ in different parameter setting. 
(a) $B=0, 600~$MeV for $m_0=500, 800~$MeV;
(b) $B=600~$MeV for different $m_{0}$. 
NJL parameters $(H, g_V)/G$ are chosen to be (1.45,1.3)$_{m_{0}=400{\rm MeV}}$, (1.6,1.3)$_{m_{0}=500{\rm MeV}}$, (1.6,1.3)$_{m_{0}=600{\rm MeV}}$, and (1.6,1.2)$_{m_{0}=700{\rm MeV}}$.
}
\label{mass-radius}
\end{figure}

In the following analysis, we fix the value $L_0=57.7\,$MeV and 
the parameter $\lambda_8' = 0$. The value of $\lambda_{10}'$ is determined as explained in sec.~\ref{sec:parameters}.
For example, $\lambda_{10}' = 0.5221$ is obtained for $m_0= 700$\,MeV below.
Then, we 
examine the effects of U(1)$_A$ anomaly on the $M$-$R$ relation. 
In Fig.~\ref{mass-radius}(a), we  show $M$-$R$ curves for several values of the anomaly strength $B$,
with the NJL parameters $(H, g_{V})$ leading to the largest and second largest maximum masses for a given set of the PDM parameters.
This shows that, due to the softening effect of anomaly as explained in sec.~\ref{sec:softening anomaly},
even the stiffest connection for $m_{0}=800~$MeV with $B=600$\,MeV is unable to satisfy the maximum mass constraints.
The effect of the anomaly in general softens EOS from low to high densities, and increasing $B$ from 0 to 600 MeV (while retuning the other parameters to reproduce nuclear saturation properties) reduces each of $M$ and $R$ by a few percent.
In Fig.~\ref{mass-radius}(b), 
we set $B=600~$MeV to fit the $\eta'$ mass and examine several values of $m_0$.
These results should be regarded as the representatives of the present review.
%
%

In this review, 
the mass of the millisecond pulsar PSR J0740+6620~\cite{Fonseca:2021wxt}
%
\begin{align}\label{cond: mass}
	M_\mathrm{TOV}^\mathrm{lowest}=2.08^{+0.07}_{-0.07}\,M_\odot \ ,
\end{align}
is regarded as the lower bound for the maximum mass, which is shown by 
upper red-shaded area in Figs.~\ref{gv_H} and \ref{mass-radius}. 
Actually, the lower bound may be even significantly higher; 
the recent analyses for the black-widow binary pulsar PSR J0952-0607 suggest the maximum mass of $2.35\pm 0.17 M_{\odot}$ \cite{Romani:2022jhd}.
Meanwhile, there are constraints, $M_{\rm max} \lesssim 2.16^{+0.17}_{-0.15} M_\odot$, 
from the gamma-ray burst GRB170817A associated with the GW170817 event
(under the assumption that the post-merger of GW170817 is a hypermassive NS).
If the maximum mass is indeed $\sim 2.3M_\odot$ or higher, 
we will need to allow much stiffer low density EOS with which much stiffer quark EOS becomes possible.
The analyses based on another criterion will be presented elsewhere.

Another important constraint comes from NS radii.
We show the constraints to the radii obtained from 
the LIGO-Virgo~\cite{LIGOScientific:2017vwq,LIGOScientific:2017ync,LIGOScientific:2018cki}
by green shaded areas on the middle 
left
\footnote{
More precisely, the LIGO-Virgo constrains the tidal deformability $\tilde{\Lambda}$ which
is the function of the tidal deformability of each neutron star ($\Lambda_1$ and $\Lambda_2$) and the mass ratio $q=M_2/M_1$.
But for EOS which do not lead to large variation of radii for $M \gtrsim 1M_\odot$, 
$\tilde{\Lambda}$ is insensitive to $q$. In fact the radii of neutron stars and $\tilde{\Lambda}$ can be strongly correlated
(for more details, see Refs.~\cite{De:2018uhw,Radice:2017lry}),
and for our purposes it is sufficient to directly use the estimates on the radii given in Ref. \cite{LIGOScientific:2018cki}, rather than $\tilde{\Lambda}$.
}
and from the NICER in Ref.~\cite{Miller:2019cac} 
by red shaded areas on the middle right. 
The inner contour of each area contains $68\%$ of the posterior probability ($1\sigma$), 
and the outer one contains $95\%$ ($2\sigma$). 
These values (plus another NICER result in Ref.~\cite{Riley:2019yda}) are summarized in Table~\ref{table:radii}. 
\begin{table}
\caption{Radius constraints for neutron stars for $\simeq 1.4M_\odot$ and $\simeq 2.1M_\odot$ NSs.  }
\begin{center}
\begin{tabular}{c|c|c}
\hline\hline
 &  radius [km] & mass [$M_\odot$] \\
 \hline
~~ GW170817 (primary) 
~~&~~     $11.9_{-1.4}^{+1.4} $     ~~&~~  $1.46_{-0.10}^{+0.12} $ ~~ \\
~~ 
~~ GW170817 (second) 
~~&~~     $11.9_{-1.4}^{+1.4} $     ~~&~~  $1.27_{-0.09}^{+0.09} $ ~~ \\

~~ J0030+0451 (NICER \cite{Miller:2019cac}) ~~&~~ $13.02_{-1.06}^{+1.24}$ ~~&~~ $1.44_{-0.14}^{+0.15} $ ~~ \\
~~ J0030+0451 (NICER \cite{Riley:2019yda}) ~~&~~ $12.71_{-1.19}^{+1.14}$ ~~&~~ $1.34_{-0.16}^{+0.15} $ ~~ \\

~~ PSR J0740+6620 (NICER \cite{Miller:2021qha}) ~~&~~ $12.35_{-0.75}^{+0.75} $ ~~&~~ $2.08_{-0.07}^{+0.07} $ ~~ \\
~~ PSR J0740+6620 (NICER \cite{Riley:2021pdl})   ~~&~~ $12.39_{-0.98}^{+1.30}$ ~~&~~  $2.08_{-0.07}^{+0.07} $ ~~ \\
\hline\hline
\end{tabular}
\end{center}
\label{table:radii}
\end{table}
From all the constraints, we restrict the chiral invariant mass as
\begin{align}
    400\, {\rm MeV} \lesssim m_{0} \lesssim 700\, {\rm MeV} \,.
\end{align}
which is updated from those in the the original work Ref.~\cite{Minamikawa:2020jfj},
$600$ MeV $\lesssim m_0 \lesssim  900$ MeV,
which corresponds to the set $\lambda'_8=\lambda'_{10}=0$ and $B=0$ in the present model.

\section{Chiral condensates in crossover}
\label{sec:CCC}

The method of interpolation can be used not only to construct a unified EOS but also to calculate microscopic quantities such as condensates and matter composition.
In hadronic and quark matter domains we consider the generating functional with external fields coupled to quantities of interest, and then interpolate two functionals.
The microscopic quantities are then extracted by differentiating the unified generating functional.
We first review the computations in hadronic and quark matter domains, and then turn to computations in the crossover region.

\subsection{Chiral condensates in the PDM}
\label{sec:CC-PDM}

The chiral condensate in the PDM can be calculated by differentiating a thermodynamic potential with respect to the current quark mass. 
In the present model the explicit chiral symmetry breaking enters only through the $V_{SB}$ term in Eq.~(\ref{VSB term}) which leads to $-\left(2 c m_{u} \sigma+c m_{s} \sigma_{s}\right)$ as in Eq.~(\ref{mean field potential}).
There may be the mass dependence in the other coupling constants in front of higher powers in meson fields, 
but such couplings exist already at $m_q=0$ and the finite $m_q$ is supposed to give only minor corrections.
Hence we neglect the $m_q$ dependence except the terms in $V_{SB}$.
Using the Gell-Mann--Oakes--Renner relation, the explicit symmetry breaking term can be written as
\begin{align}
\Omega_{\rm ESB} = - \left(2 c m_{u} \sigma+c m_{s} \sigma_{s}\right) = m_q \langle ( \bar{u} u + \bar{d} d )  \rangle_0 \frac{\, \sigma  \,}{\, f_\pi \,} + m_s \langle \bar{s} s \rangle_0 \frac{\sigma_s}{\sigma_{s0}} \,,
\end{align}
where 
$ \langle (\bar{u}u +\bar{d} d ) \rangle_0$ and $\langle \bar{s} s \rangle_0$ are the chiral condensates in vacuum. 
The in-medium chiral condensates are obtained as
\begin{align}
& \langle ( \bar{u} u + \bar{d} d ) \rangle 
\equiv \pdv{\Omega_{\rm ESB}}{m_q}
=  \langle ( \bar{u} u + \bar{d} d ) \rangle_0 \frac{\, \sigma  \,}{\, f_\pi \,} \,, \\
& \langle \bar{s} s \rangle \equiv \pdv{\Omega_{\rm ESB}}{m_s}
=  \langle  \bar{s} s \rangle_0 \frac{\, \sigma_s  \,}{\, \sigma_{s0} \,} \,,
\end{align}
where we neglected 
$m_q$ and $m_s$ dependences of $\langle ( \bar{u} u + \bar{d} d ) \rangle_0$ and $\langle \bar{s} s \rangle_0$ which are of higher orders in $m_q/M_q$ and $m_s/M_s$.

In the following sub-subsection~\ref{sec:CSD}, we examine how $\sigma$ varies as baryon density increases,
and 
we study the 
in-medium $\langle ( \bar{u} u + \bar{d} d ) \rangle$ condensate in sub-subsection~\ref{sec:dilute}.
We postpone discussions on the strange quark condensate $\langle \bar{s} s \rangle$ to 
subsection~\ref{sec:unified_condensate}
since changes in $\langle \bar{s} s \rangle$ at $n_B \le 2n_0$, which are induced only through the anomaly, are very small in the hadronic region.

\subsubsection{Chiral scalar density in a nucleon}\label{sec:CSD}

To set up the baseline for the estimate of in-medium chiral condensates, 
we consider the scalar charge, $N_\sigma $, for a nucleon in vacuum. 
It is defined as 
\begin{align}
N_\sigma 
&=
\int_{  \mathbf{x} }
\langle N| ( \bar{u} u + \bar{d} d ) (x) | N \rangle 
= \langle N | \frac{\, \partial  H_{\rm QCD} \,}{\, \partial m_q \,}  |N \rangle 
= \frac{\, \partial m_N^{\rm vac} \,}{\, \partial m_q \,}  \,,
\end{align}
where $H_{\rm QCD}$ is the QCD Hamiltonian. 
In the last step we used the Hellmann--Feynman theorem~\cite{Gubler:2018ctz}.

In the PDM,  the current quark masses affect nucleon masses only through the modification of $\sigma$.
The nucleon's chiral scalar charge at vacuum is given as
\begin{align}
N_\sigma
 \equiv \frac{\, \partial m^{\rm vac}_N \,}{\, \partial m_q \,} 
 = \frac{\, \partial \sigma_0 \,}{\, \partial m_q \,} \bigg( \frac{\, \partial m_N \,}{\, \partial \sigma \,} \bigg)_{\sigma = \sigma_0} \,.
 \label{eq:Nsigma}
\end{align}
The mass derivative of $\sigma_0$ is related to the chiral susceptibility which is given by
the (connected) scalar correlator at zero momentum,
\begin{align}
\frac{\, \partial \langle \bar{q} q (x) \rangle \,}{\, \partial m_q \,}
&\sim 
\int {\cal D} q {\cal D} \bar{q}
~ [ \bar{q} q (x) ] \frac{ \partial }{\, \partial m_q \,} \bigg( {\rm e}^{ - \int_{x'} m_q \bar{q} q (x') + ...} /Z \bigg)
\nonumber \\
&\sim 
\int_{x'} \big\langle [ \bar{q} q (x) ] [ \bar{q} q (x') ] \big\rangle_{\rm conn.}
\sim
 \lim_{q\rightarrow 0} \frac{1}{\, q^2 + m_\sigma^2 \,} \,.
\end{align}
Then, a smaller scalar meson mass enhances $N_\sigma$.

Multiplication of $m_q$ to the scalar charge leads to the so-called nucleon sigma term:
\begin{align}
\Sigma_N \equiv m_q N_\sigma = \int_{\mathbf{x}} \langle N| m_q (  \bar{u} u + \bar{d} d ) | N \rangle \,,
\end{align}
which is renormalization group invariant, and has direct access to experimental quantities.
The traditional estimate~\cite{Gasser:1990ce}
gives $\Sigma_N \simeq 45$ MeV. 
But the precise determination is difficult, 
and the possible range is $40\text{--}70$ MeV,
according to lattice QCD analyses or combined analyses of the lattice QCD 
and the chiral perturbation theory.
(See Ref. \cite{Gubler:2018ctz} for a review and the references therein.)
Here we take $m_q\simeq 5$\,MeV which leads to obtain $N_\sigma \simeq 8\text{--}14$, and the scalar density is given by
\begin{align}
 \frac{N_\sigma}{\frac{4}{3}\pi R_{N}^3}
 =\bigg( 0.24\text{--}0.30\, {\rm GeV}\times\frac{\, 1\,{\rm fm}\,}{\, R_N \,} \bigg)^3\,,
\end{align}
where $R_N \sim 1\,$fm is the size of a nucleon.
(Note that the scalar isoscalar radius is estimated as 
$\langle r_s^2 \rangle \simeq ( 0.7\text{--}1.2\, {\rm fm})^2$~\cite{Schweitzer:2003sb}.)
Note that the magnitude is roughly the same order as the vacuum one, 
but the sign is opposite.
Therefore the nucleon scalar charges tend to cancel the vacuum one and reduce the net value of $\sigma $.
Therefore the appearance of nucleons inevitably reduces the magnitude of chiral condensates.

Table~\ref{tab-sigmatermvalue}
summarizes the $\sigma$-dependence of the nucleon mass ($\partial m_N/\partial \sigma$), 
the scalar meson mass ($m_\sigma$), and the nucleon sigma term ($\Sigma_N$) predicted 
by the PDM for several choices of $m_0$.
\begin{table}\centering
\caption{ 
$\sigma$-dependence of the nucleon mass, the scalar meson mass, and the nucleon sigma term predicted by the PDM in vacuum. 
}\label{tab-sigmatermvalue}
	\begin{tabular}{c|cccc}
		\hline\hline
		$m_0$ [MeV] ~&~ 400 ~&~ 500 ~&~ 600 ~&~ 700 ~  \\
		\hline
$(\partial m_N /\partial\sigma)_\mathrm{vac.}$ & 8.79 &~ 7.97 ~&~ 7.01 ~&~ 5.87 ~ \\
$m_\sigma$ [MeV]   &607    &   664 &       688 &        599 \\
$\Sigma_N$ [MeV]   & 51.12 &     48.71 &    51.39 &    62.01 \\
		\hline\hline
	\end{tabular}
\end{table}
The estimates of $m_\sigma$ and $\Sigma_N$ 
are reasonably consistent with the hadron phenomenology;
 $m_\sigma$ are consistent with the mass of the scalar meson $f_0(500)$ 
(with the width $\sim 500 $ MeV)\footnote{
It is not a trivial issue whether one can identify $\sigma$ in mean field models with the physical scalar meson.
},
and the estimates on the nucleon sigma term, $\Sigma_N \simeq 40 $-$70$\,MeV, are within the ball park of several theoretical estimates.

\subsubsection{Dilute regime}\label{sec:dilute}

\begin{figure}\centering
\includegraphics[width=0.4\hsize]{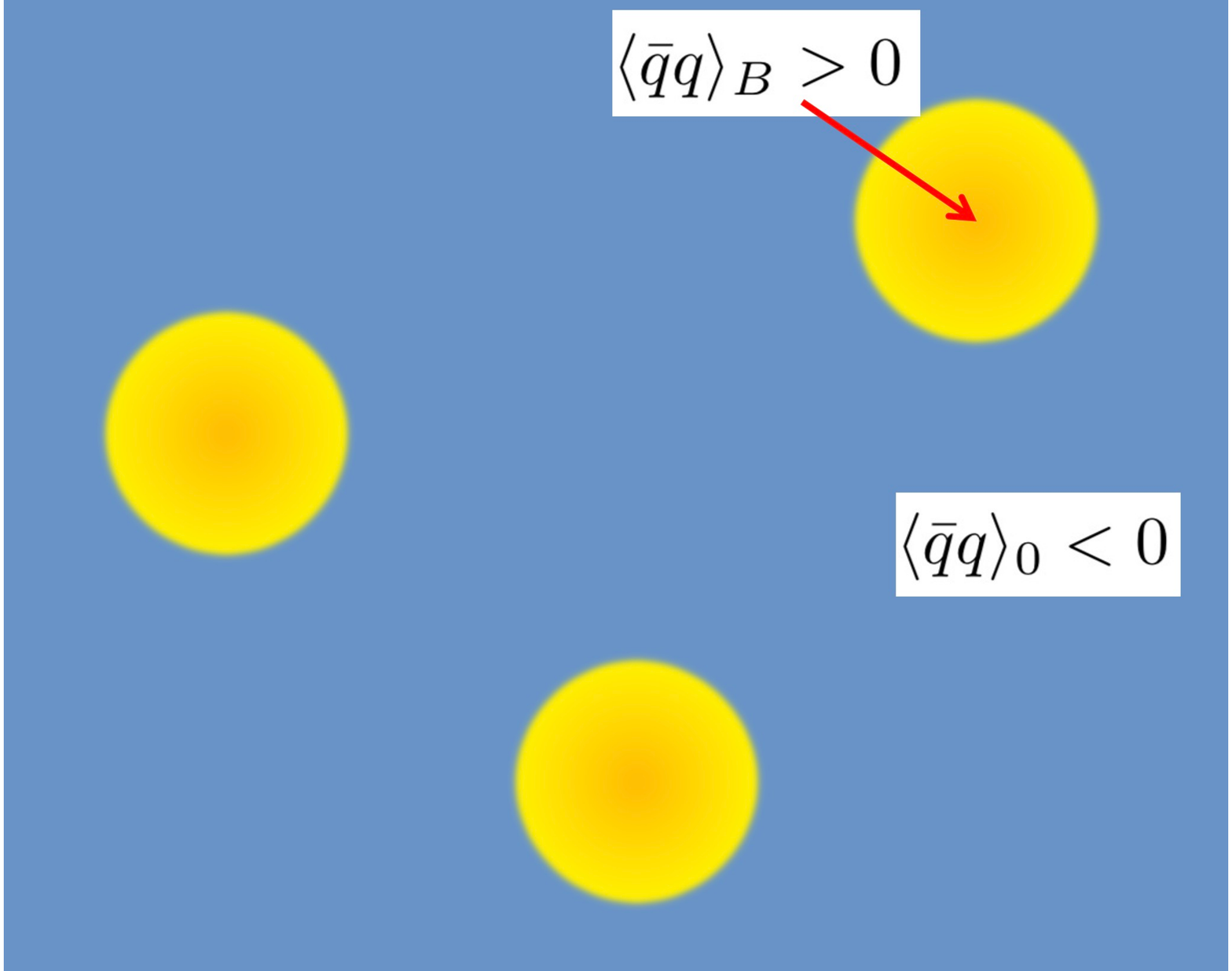}
\caption{
Schematic picture of the chiral condensates in dilute regime. 
The chiral scalar charge is negative where the vacuum chiral condensate dominates,
while nucleons contribute to the positive scalar charges to cancel the vacuum contributions.
}
\label{fig:chiral_in_nuclear}
\end{figure}
In the dilute regime (Fig.~\ref{fig:chiral_in_nuclear}), nucleons are widely separated,
In good approximation the in-medium scalar density is simply the sum of negative scalar charges from chiral condensates and the positive scalar charges from nucleons
(linear density approximation  (LDA)),
\begin{align}
\langle (  \bar{u} u + \bar{d} d ) \rangle \simeq \langle (  \bar{u} u + \bar{d} d ) \rangle_0 + n_B N_\sigma \,, 
\end{align}
which can be rewritten as
\begin{align}
\sigma \simeq f_\pi \bigg( 1 + n_B  \frac{\, N_\sigma \,}{\, \langle (  \bar{u} u + \bar{d} d ) \rangle_0 \,} \bigg) \,.
\end{align}
In this LDA, the $\sigma$ decreases linearly as a function of $n_B$.

\begin{figure}\centering
\includegraphics[width=0.6\hsize]{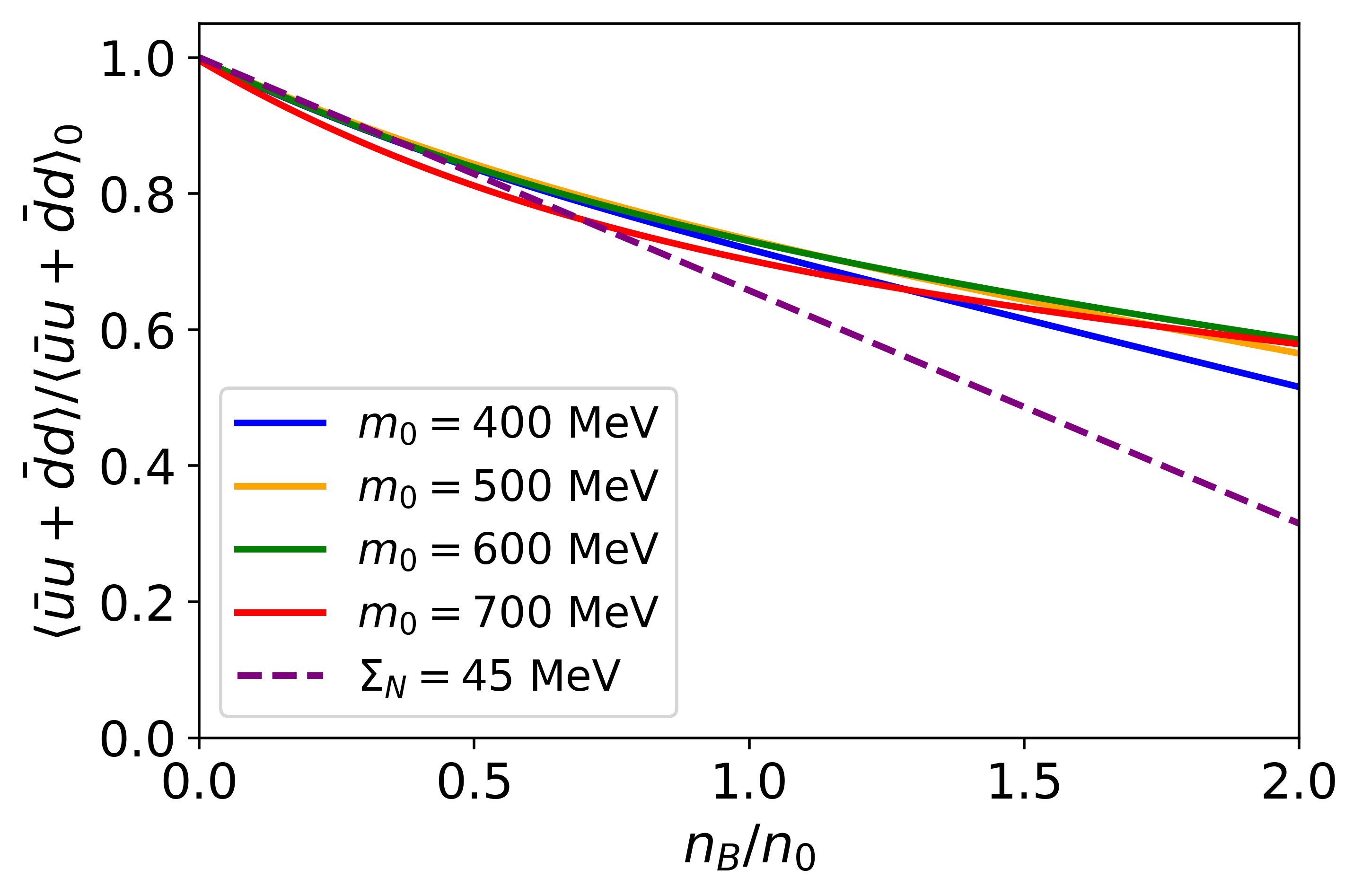}
\caption{{
Dependence of the quark condensate in the PDM 
$\ev{\bar uu}/\ev{\bar uu}_0 = \sigma/f_\pi$ on the neutron number density $n_n$ for $m_0=$ 400, 500, 600, and 700 MeV. 
Here the condensate is normalized by the vacuum counter part.
}}
\label{fig-condensate}
\end{figure}

The linear density approximation is violated when density increases and nonlinear effects set in.
Shown in Fig.~\ref{fig-condensate} are the ratio of the quark condensate,
$\ev{\bar uu}/\ev{\bar uu}_0 = \sigma/f_\pi$, as a function of the neutron number density $n_n$ in pure neutron matter. 
The result of the linear density approximation is also shown for comparisons.
Our mean field results 
are consistent with the linear density approximation with $\Sigma_N = 45\,$MeV in the low-density region.
Our predictions start to deviate from the LDA around $n_B = 0.5 n_0$, signaling the importance of higher powers of $n_B$.

\begin{figure}\centering
\includegraphics[width=0.6\hsize]{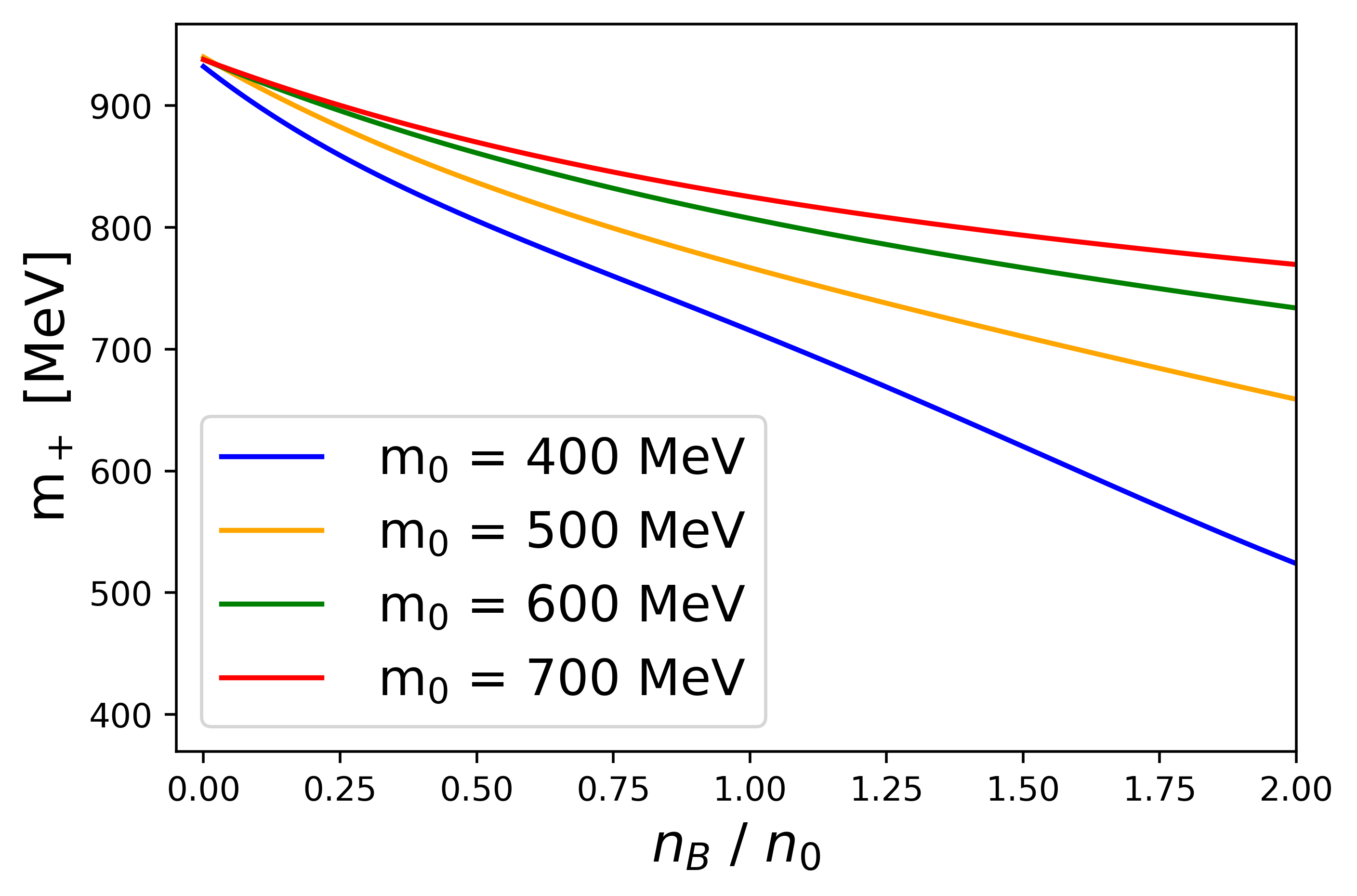}
\caption{{
Dependence of the nucleon masses in the PDM on the neutron number density $n_n$, for $m_0=$ 400, 500, 600, and 700 MeV. 
}}
\label{fig-mN}
\end{figure}

We stress that, in the PDM, 
while the chiral restoration or reduction of $\sigma$ occurs rather quickly with increasing density, 
such changes do not immediately mean the structural changes in nucleons nor in the nucleon or quark Dirac sea. 
The nucleon mass in the PDM is relatively modest (Fig.~\ref{fig-mN}), 
and this feature is welcomed for commonly used no sea approximation for the thermodynamic potential (see Eq.(\ref{PDM: grand functional}))
which is justified only when modifications in the Dirac sea are small.
Another hint on the chiral condensates and hadron structures 
comes from a high temperature transition where a hadron resonance gas (HRG) transforms to a quark gluon plasma (QGP).
There, the chiral condensates begin to drop before the temperature reaches the critical temperature, 
but the HRG model with the {\it vacuum} hadron masses remains valid in reproducing 
the lattice data even after chiral condensates are substantially reduced~\cite{Karsch:2003zq,Andronic:2017pug}. 
The chiral restoration beyond cancellations of negative and positive scalar charges will be discussed in the next section for quark matter models.
%

\subsection{Chiral condensates in the CFL quark matter}

\begin{figure}\centering
\includegraphics[width=0.5\hsize]{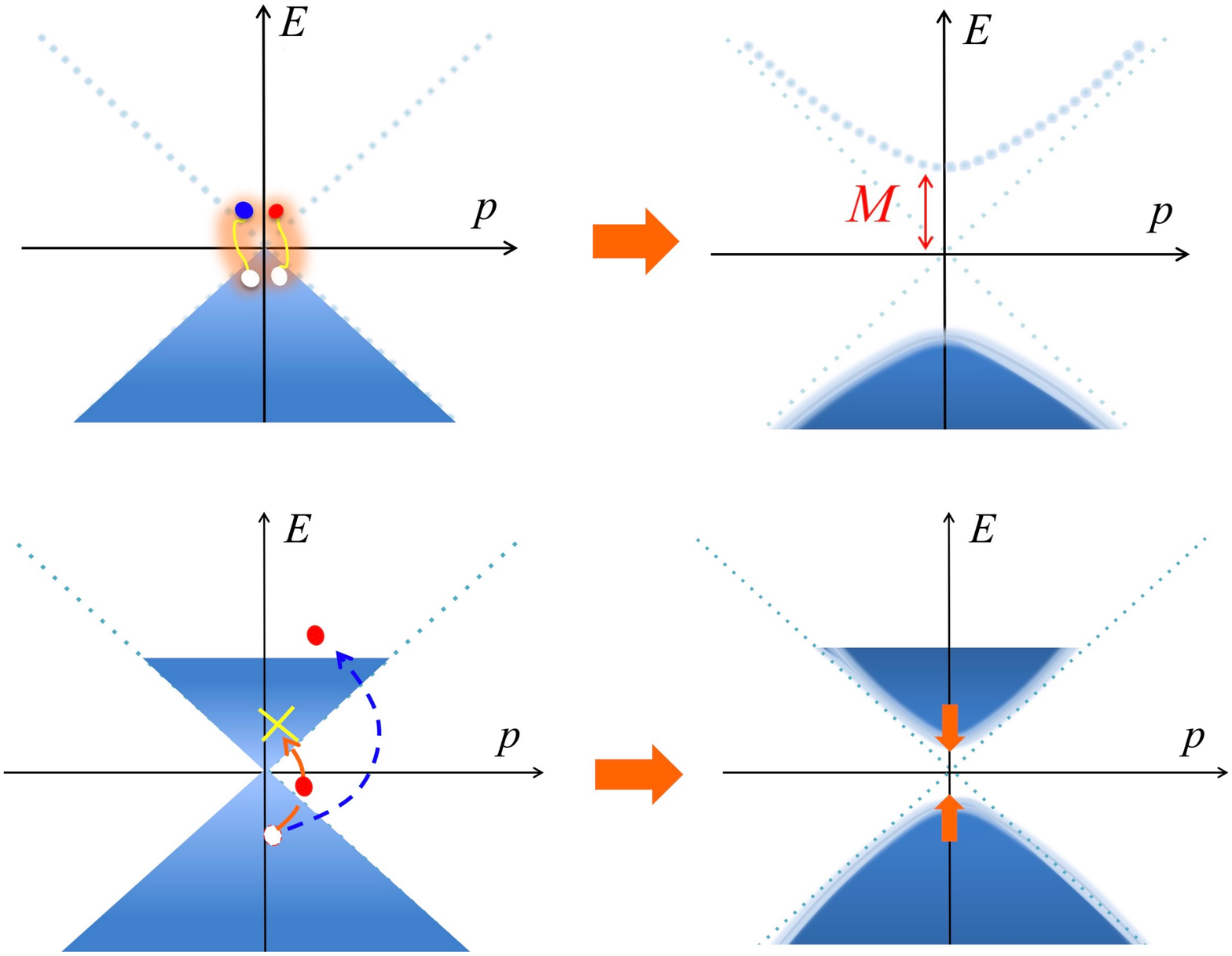}
\caption{{
Chiral symmetry breaking by condensation of quark-antiquark pairs; (upper) in vacuum; (lower) in medium. In the latter the pairing is blocked by the quark Fermi sea.
}}
\label{fig:chiral_in_quark}
\end{figure}

In terms of quarks, the chiral condensates are triggered by the attractive quark-antiquark pairing.
At high density, such pairing is disfavored by the presence of the quark Fermi sea;
as shown in 
Fig.~\ref{fig:chiral_in_quark},  
creating an antiquark costs about the quark Fermi energy
since 
it is necessary to bring a particle in the Dirac sea to the domain beyond the Fermi sea. 
Therefore, the chiral condensates made of quarks and antiquarks naturally dissociate as density increases. 
Instead, the particle-particle \cite{Alford:2007xm} 
or particle-hole pairings~\cite{Shuster:1999tn,Kojo:2009ha,Kojo:2011cn} near the Fermi surface
do not have such energetic disadvantages.
The method of computations is given in Sec. \ref{sec:QM NJL}.

We note that, unlike the chiral restoration in dilute nuclear matter as a mere consequence of cancellations of positive and negative charges,
in quark matter the magnitude of each contributions is reduced, together with the chiral restoration in the quark Dirac sea. 
This extra energy from the Dirac sea modification is important in quark matter EOS and must be taken into account.
The softening of quark EOS due to the U(1)$_A$ anomaly is related to the Dirac sea modifications associated with 
the chiral restoration\footnote{This statement needs qualification. 
If we consider the anomaly term for couplings between diquark and chiral condensates \cite{Hatsuda:2006ps}, 
then the EOS can be stiffer, see Fig.~7 in Ref.\cite{Kojo:2014rca}.
}.

\subsection{Condensates in a unified EOS }
\label{sec:unified_condensate}

In this subsection, we review the interpolating method of generating functionals which is introduced in
Ref.~\cite{Minamikawa:2021fln}.
We use it to calculate the chiral and diquark condensates from nuclear to quark matter,
and also to examine the composition of matter with ($u,d,s$)-quarks and charged leptons (electrons and muons, $e,\mu$).

\subsubsection{Unified generating functional}

For computations of condensates $\phi$, 
we first construct a generating functional $P(\mu_B; J)$ with external fields $J$ coupled to the $\phi$. 
A condensate $\phi$ at a given $\mu_B$ is obtained by 
differentiating $P(\mu_B; J)$ with respect to $J$ and then set $J=0$,
\begin{align}
\phi = - \frac{\, \partial P \,}{\, \partial J \,} \bigg|_{J=0} \,.
\end{align}
The generating functional for the nuclear domain, $n_B \leq 2n_0$, is given by the PDM, and for the quark matter domain, $n_B \geq 5n_0$, by the NJL-type model.
We interpolate these functionals with the constraints that  the interpolating curves match up to the second derivatives at each boundary, $2n_0$ and $5n_0$.
For the interpolating function, we adopt a polynomial function of $\mu_B$ with six coefficients $a_n(J)$,
\begin{align}
P_\interp(\mu_B ; J)&=\sum_{n=0}^5a_n(J)\mu_B^n\,.
\label{eq:unified_functional}
\end{align}
%
We determine the chemical potentials at the boundaries, $\mu_B^L$ and $\mu_B^U$, as 
\begin{align}
n_B (\mu_B^L; J) = 2n_0 \,, \quad n_B (\mu_B^U; J) = 5n_0 \,.
\end{align}
The resulting $\mu_B^L$ and $\mu_B^U$ depend on $J$.
The six boundary conditions
\begin{align}
\frac{\, \partial^k P_\interp \,}{\, ( \partial \mu_B )^k \,} \bigg|_{\mu_B^L (\mu_B^U) }
	= \frac{\, \partial^k P_{\rm PDM (NJL)} \,}{\, ( \partial \mu_B )^k \,} \bigg|_{\mu_B^L (\mu_B^U) } \,,
\label{eq:matching}
\end{align}
with $k=0,1,2$ uniquely fix $a_n$'s. 
As in the EOS construction, the generating functional must satisfy the causality condition.
Such constraints are transferred to the evaluation of condensates;
condensates in the crossover domain are correlated with those in nuclear and quark matter.

\subsubsection{An efficient method for computations of many condensates}

While the generating functional in the previous section is general, 
the calculations become cumbersome when we need to compute many condensates.
Each condensate requires the corresponding external field and generating functional.
Fortunately, for the interpolating function Eq.(\ref{eq:unified_functional}), 
we can use a more efficient method in Ref.~\cite{Minamikawa:2021fln}
which does not demand construction of $P(\mu_B,J)$ 
but utilizes only the $\mu_B$-dependence of the condensate at $J=0$ for each interpolating boundary.

In the interpolated domain the condensate $\phi$ can be expressed as
\begin{align}
\phi_\interp =-\pdv{P_\interp }{J}\bigg|_{J=0}=-\sum_{n=0}^5{\pdv{a_n}{J}}\bigg|_{J=0}\mu_B^n\,. 
\label{eq:inter_condensate}
\end{align}
This implies the equivalence between the determination of $\phi_\interp$ and that 
of six constants $\partial a_n/\partial J \big|_{J=0}$.
Taking the $\mu_B$-derivatives in Eq.(\ref{eq:matching}), we obtain
\begin{align} \label{eq-dphidmu}
\frac{\partial}{\, \partial J \,} \bigg( \frac{\, \partial^k P_\interp \,}{\, ( \partial \mu_B )^k \,} \bigg|_{\mu_B^L (\mu_B^U) } \bigg)
	= \frac{\partial}{\, \partial J \,} \bigg( \frac{\, \partial^k P_{\rm PDM (NJL)} \,}{\, ( \partial \mu_B )^k \,} \bigg|_{\mu_B^L (\mu_B^U) } \bigg) \,,
\end{align}
where $k=0,1,2$. 
Only quantities at a given $\mu_B$ and $J=0$ are necessary to construct all these derivatives at $J=0$.
Hence this method speeds up our analyses considerably.

\subsubsection{Numerical results}\label{sec:numerical results}

Using the method explained above,
we calculate the light quark chiral condensate $\ev{ (\bar{u}u + \bar{d} d) }$, 
the strange quark condensate $\ev{\bar{s}s}$, 
the diquark gaps $\Delta_j$ ($j=1,2,3$), and the quark number densities $n_f$ ($f=u,d,s$),
from nuclear to quark matter domain. 
Below, we adopt three values of the chiral invariant mass ($m_0 = 500$, $600$, $700$\,MeV) as samples
with fixing the anomaly coefficient $B$ to $600$\,MeV and the NJL parameters $(H/G, g_V/G)$ to $(1.45,0.5)$.
%
%
The presence of the anomaly term in the PDM is the difference between the results in this review and in Ref.~\cite{Minamikawa:2021fln}
whose impacts are just few percents in magnitude.
The EOS for these parameter sets satisfies $ 0 \le c_s^2 \le 1$.
For comparisons, the extrapolation of the PDM results are shown by black dotted curves.

\paragraph{\bf Light quark chiral condensates}\label{sec:light_quark_condensate}

\begin{figure}\centering
\includegraphics[width=0.5\hsize]{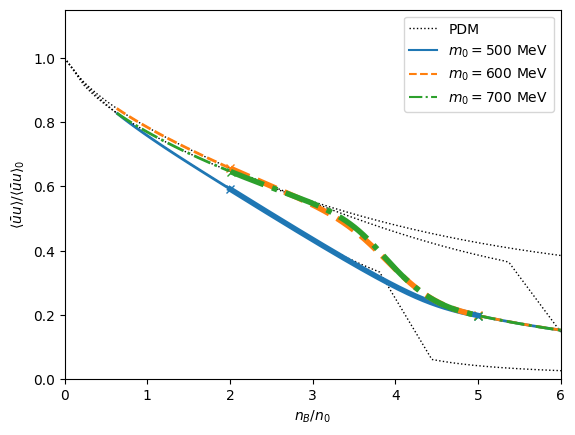}
\caption{{
Density dependence of the chiral condensates normalized by the vacuum counterpart.
 The parameters are chosen as $B=600$\,MeV and $(H/G, g_V/G) = (1.45,0.5)$. 
}}
\label{fig-interpolate-qqbar}
\end{figure}
Figure~\ref{fig-interpolate-qqbar} shows the density dependences of the in-medium chiral condensate normalized by the vacuum value, 
$\ev{ (\bar{u}u + \bar{d} d)} /\ev{ (\bar{u}u + \bar{d} d) }_0$.
Clearly, the condensate at the boundaries affects the condensate in the crossover region.

The condensates in the hadronic matter strongly depend on
the choice of $m_0$:
for $m_0 =500$ MeV, the nucleon mass $m_N = 939$ MeV gains a large contribution from the chiral condensate,
and the Yukawa coupling of nucleons to $\sigma$ is large;
accordingly the chiral condensate drops quickly as baryon density increases.
For a larger $m_0$, the nucleons have less impacts on the chiral condensates,
and the chiral restoration takes place more slowly.

As mentioned in Sec.~\ref{sec:CC-PDM}, 
the PDM may underestimate chiral restoration effects as they do not describe the chiral restoration at quark level.
Putting quark matter constraints from the high density and using the causality constraints for the interpolated domain,
we can gain qualitative insights how the chiral restoration should occur toward high density.
Taking into account nuclear and quark matter effects, the interpolation offers reasonable descriptions for the crossover domain.

\paragraph{\bf Strange chiral condensates} 


\begin{figure}\centering
\includegraphics[width=0.5\hsize]{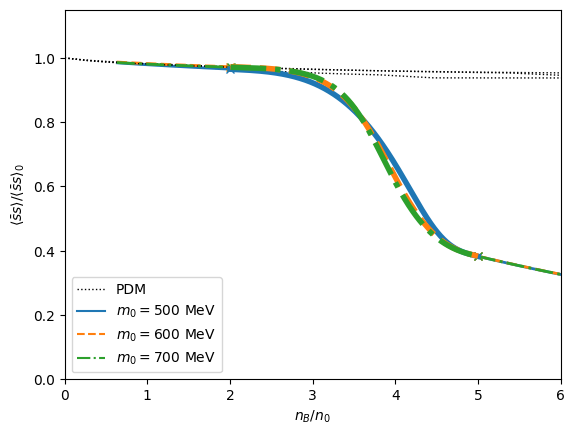}
\caption{{
Density dependence of the strange quark condensate normalized by the vacuum counterpart. 
The parameters $B$ and $(H/G,g_V/G)$ are as in Fig.~\ref{fig-interpolate-qqbar}. 
}}
\label{fig-interpolate-ssbar}
\end{figure}
The density dependence of the strange quark condensate is shown in Fig.\ref{fig-interpolate-ssbar}.
In the present PDM model, the $\sigma_s$ field corresponding to the strange quark condensate does not directly couple to nucleons, 
but only through the anomaly term in the meson potential, Eq.~(\ref{V anom}).
As a result, the density dependence is mild in the hadronic matter ($n_B \leq 2 n_0$).
In the interpolated region, the condensate starts to decrease rapidly toward the one in the quark matter which is about $40\%$ of the vacuum value at $n_B = 5 n_0$.
There are at least two effects responsible for this chiral restoration. 
The first is the reduction of the anomaly contribution, $\sim \langle \bar{u}u \rangle \langle \bar{d}d \rangle (\bar{s}s)$,
which is due to the chiral restoration for light quark sectors.
The other is due to the evolution of the strange quark Fermi sea.
In our unified model the strangeness sector significantly deviates from the prediction of the PDM at $n_B \simeq 3n_0$, 
due to the constraints from the quark matter boundary conditions.

\paragraph{\bf Diquark gaps and number density} 

\begin{figure}\centering
	\includegraphics[width=0.49\hsize]{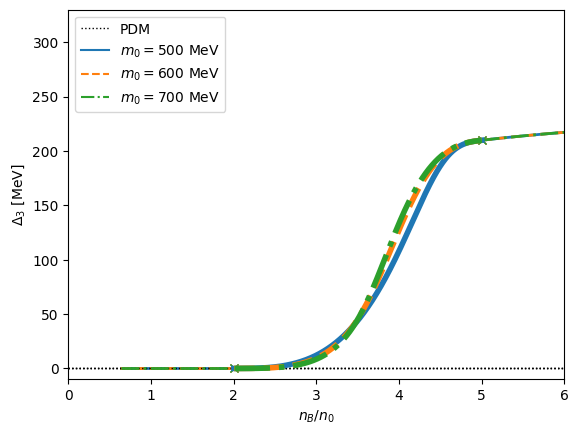}
	\includegraphics[width=0.49\hsize]{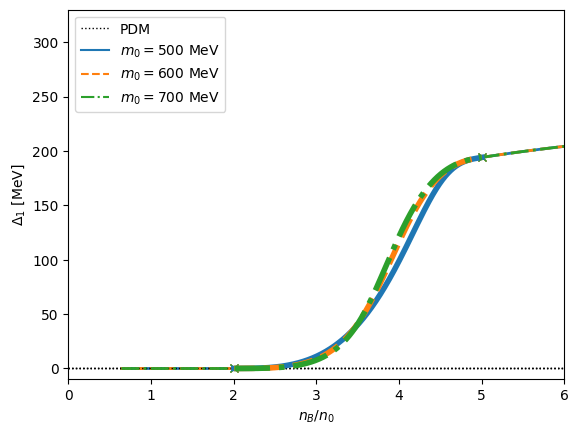}
\caption{{ 
Diquark condensates as functions of density; (left) $\Delta_3 = \Delta_{ud}$, and (right) $ \Delta_1 = \Delta_{ds}$.
The parameters $B$ and $(H/G,g_V/G)$ are as in Fig.~\ref{fig-interpolate-qqbar}. 
}}
\label{fig:diquarks}
\end{figure}
Shown in Figure~\ref{fig:diquarks} are the diquark gaps in the $ud$-pairing channel (left panel),  $ds$-pairing channel (right panel) at various densities.
We set the diquark condensates to zero at $n_B \le 2n_0$. 
Meanwhile the isospin symmetry holds in the CFL quark matter,
so in the whole region $\Delta_{ds} \simeq \Delta_{us}$ holds in good accuracy.

\begin{figure}\centering
	\includegraphics[width=0.49\hsize]{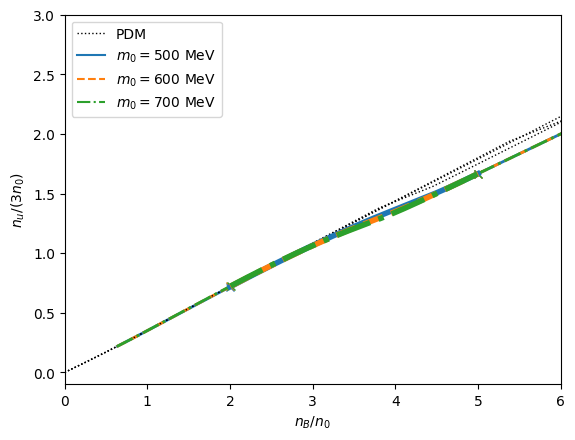}
	\includegraphics[width=0.49\hsize]{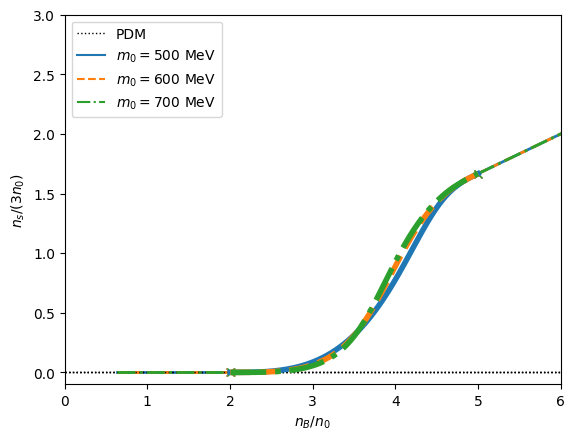}
\caption{{ 
Number density for (left) up-quarks, and (right) strange-quarks.
The parameters $B$ and $(H/G,g_V/G)$ are as in Fig.~\ref{fig-interpolate-qqbar}. 
}}
\label{fig:number}
\end{figure}
Next we study the density dependence of diquark condensates on quark number density (Fig.\ref{fig:number}). 
The quark densities in nuclear domain are calculated as $n_u=2n_p+n_n$, $n_d=n_p+2n_n$, and $n_s=0$. 
As seen from Figs.~\ref{fig:diquarks} and \ref{fig:number}, 
there are clear correlations between the growth of the diquark condensates and of quark number densities.
These two quantities assist each other:
more diquark pairs are possible for a larger quark Fermi sea,
while the resulting energy reduction in turn enhances the quark density.
The flavor composition is also affected by these correlations:
the substantial $u,d$-quark Fermi sea and the pairing to strange quarks favor the formation of the strange quark Fermi sea,
even before the quark chemical potential reaches the threshold of the vacuum strange quark mass.

\paragraph{\bf Quark and lepton compositions}

\begin{figure}\centering
	\includegraphics[width=0.5\hsize]{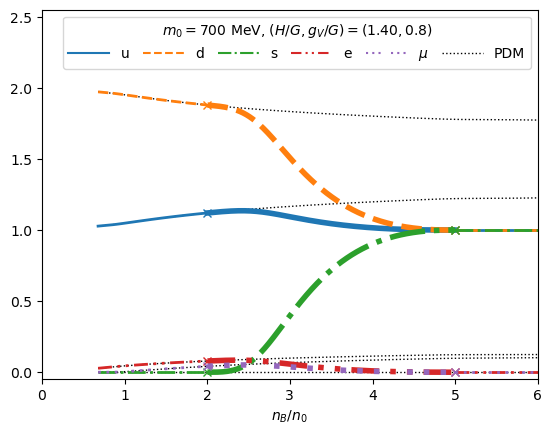}
\caption{{ 
Matter composition $n_f/n_B$ ($f=u,d,s$) and $n_l/n_B$ as functions of baryon density.
}}
\label{fig:udsl}
\end{figure}
The present framework can be extended for computations of the matter composition in NS matter with leptons.
We just impose the charge neutrality and $\beta$-equilibrium conditions on the generating functionals.
Shown in figure \ref{fig:udsl} are quark flavor $n_f/n_B$ with $f=u,d,s$, and lepton fraction, $n_l/n_B$ with $l$ referring to electrons or muons.

The quick evolution of the strangeness fraction, taking off around $n_B\simeq 2.5n_0$ and becoming as abundant as up- and down-quarks at $n_B \gtrsim 4.5n_0$,
and the associated reduction of lepton fractions, are one of distinct features of our unified model.
Beyond $5n_0$, in the CFL quark matter the charge neutrality is satisfied by quarks, and no charged leptons are necessary.


\section{A summary}
\label{sec:summary}

In this review article, we summarized main points of Refs.~\cite{Minamikawa:2020jfj,Minamikawa:2021fln,Gao:2022klm} and have updated some of analyses including the U(1)$_A$ anomaly effects.
In Sec.~\ref{sec:PDM}, we explained how to construct the EOS in hadronic matter for $n_B \leq 2 n_0$ using an effective hadron model based on the parity doublet structure. 
In the analysis, we focused on the effect of U(1)$_A$ axial anomaly included as the KMT-like interaction among scalar and pseudoscalar mesons, and showed that the effect makes the EOS softer.
In Sec.~\ref{sec:QM NJL}, following Ref.~\cite{Baym:2019iky}, we briefly review how to construct a quark matter EOS for $n_B \geq 5n_0$ using a NJL-type model.
Then in Sec.\ref{sec:interpolation}, we built up a unified EOS in the density region of $2 n_0 \leq n_B \leq 5 n_0$ by interpolating hadronic and quark EOS.
For given microscopic parameters, we calculated $M$-$R$ relations of NSs, confronted them with the observational constraints, 
and then obtained constraints on the chiral invariant mass and quark model parameters.
In Sec.~\ref{sec:CCC} we determined the density dependence of the chiral condensate in the interpolated region using a method proposed in Refs.~\cite{Minamikawa:2021fln}.
The boundary conditions from the hadronic and quark matter affect condensates in the intermediate region and give a balanced description.

We would like to stress that our method provides some connection from 
microscopic physical quantities such as the chiral invariant mass, the chiral condensates and diquark gaps
to macroscopic observables such as masses and radii of NSs.
Actually, our analysis implies that rapid decrease of the nucleon mass even near the normal nuclear density, 
which can occur when the chiral invariant mass $m_0$ is very small, provides too soft EOS to satisfy the radius constraint of NSs with mass of about $1.4 M_{\odot}$.
In other words, radius constraint of NSs obtained from recent observations indicates that 
the nucleon mass should include a certain amount of chiral invariant mass, 
from which the nucleon keeps a large portion of its mass even in the high density region where the chiral symmetry restoration is expected to occur.

Our density dependence of the chiral condensate in the 
low density region is consistent with the linear density approximation.
We should note the reduction of the chiral condensate there is achieved by the contribution of the positive scalar charge of nucleon without changing the nucleon properties drastically.
This is due to our construction of hadronic matter in the PDM: We adopted so called ``no-sea approximation'' where we neglect the effect of nucleon Dirac sea and use fixed nucleon-meson couplings for $n_B \lesssim 2n_0$.
In the present treatment, the intrinsic properties of nucleons start to change at $n_B \gtrsim 2n_0$, drastically, 
where quark exchanges among baryons become frequent; since baryons are made of quarks, the quark exchanges are supposed to change the baryon structure.
Such intrinsic dependence would be able to be included through the introduction of the density (and/or temperature) dependent coupling constants in effective hadronic models as done in, e.g., Refs.~\cite{Harada:2001it,Harada:2001qz}.
This is refection of partially released quarks which are affected by the medium.
The inclusion of such effects into coupling constants is very difficult.
Our interpolation scheme provides a practical way to implement some restrictions through the quark matter constraints at high density.

In the present model for hadronic matter, we did not explicitly include the hyperons assuming that they are not populated in the low-density region $n_B \lesssim 2 n_0$.
The hyperons may enter into matter around $n_B \sim 2$-$3n_0$, which is not far from present choice for the hadronic boundary.
It would be interesting to make analysis explicitly including hyperons based on the parity doublet structure (see, e.g., Refs.~\cite{Nishihara:2015fka,Minamikawa:prep}).

In the present analysis we assume that
anomaly has stronger impact in mesonic sectors  than in baryonic sectors and included the anomaly $B$ term only in the mesonic sector.
It would be interesting to include some Yukawa interactions which also break the U(1)$_A$ symmetry.

\begin{acknowledgments}

The work of T.M., B.G., and M.H. was supported in part by JSPS KAKENHI Grant No. 20K03927. T.M. was also supported by JST SPRING, Grant No. JPMJSP2125; 
T.K. by the Graduate Program on Physics for the Universe (GPPU) at Tohoku university.

\end{acknowledgments}



\bibliography{ref_2022_review.bib}

\end{document}